\documentclass[onecolumn]{aastex701}
\usepackage{multirow}

\newcommand\iroman{{\it Roman }}
\newcommand\iromann{{\it Roman}}
\newcommand\synthpop{{\sc SynthPop }}
\newcommand\synthpopp{{\sc SynthPop}}
\defcitealias{synthpopI}{J. Kl{\"u}ter \& M. J. Huston et al.}

\received{April 9, 2026}
\revised{\today}
\submitjournal{AJ}

\newcommand\lsu{\affiliation{Department of Physics \& Astronomy, Louisiana State University, Baton Rouge, LA 70803, USA}}
\newcommand\osu{\affiliation{Department of Astronomy, The Ohio State University, Columbus, OH 43210, USA}}
\newcommand\berkeley{\affiliation{Department of Astronomy, University of California Berkeley, Berkeley, CA 94720, USA}}
\newcommand\umd{\affiliation{Department of Astronomy, University of Maryland, College Park, MD 20742, USA}}

\newcommand\vanderbilt{\affiliation{Department of Physics and Astronomy, Vanderbilt University, Nashville, TN 37235, USA}}
\newcommand\ipac{\affiliation{IPAC, Caltech, 1200 E. California Blvd., Pasadena, CA 91125, USA}}

\newcommand\carnegie{\affiliation{Observatories of the Carnegie Institution for Science, Pasadena, CA 91101, USA}}
\newcommand\jpl{\affiliation{NASA Jet Propulsion Laboratory, Pasadena, CA, 91109, USA}}
\newcommand\manchester{\affiliation{Department of Physics and Astronomy, University of Manchester, Oxford Rd, Manchester M13 9PL, UK}}

\newcommand{\llnl}{Space Science Institute, Lawrence Livermore National Laboratory, 7000 East Ave., Livermore, CA 94550, USA}

\shorttitle{Updated SynthPop Model}
\shortauthors{Huston et al.}
\graphicspath{{./}{figures/}}

\begin{document}

\title{An Updated \synthpop Model for Microlensing Simulations I: Model Description \& Evaluation
}

\author[0000-0003-4591-3201]{Macy J. Huston}
\berkeley 
\email[show]{mhuston@berkeley.edu}

\author[0000-0003-4310-3440]{Alison L. Crisp}
\osu
\email{crisp.92@osu.edu}

\author[0009-0002-1973-5229]{Marz Newman}
\lsu
\email{mnewm25@lsu.edu}

\author[0009-0009-1267-8445]{Riley Patlak}
\berkeley
\email{rileypatlak@berkeley.edu}

\author[0000-0001-7506-5640]{Matthew T. Penny}
\lsu
\email{penny1@lsu.edu}

\author[0000-0002-3469-5133]{Jonas Kl\"uter}
\lsu
\email{jonas.klueter@web.de}


\author[0000-0001-9397-4768]{Samson A. Johnson} 
\osu \jpl
\email{samson.a.johnson@gmail.com}

\author[0000-0002-1052-6749]{Peter McGill}
\affiliation{\llnl}
\email{mcgill5@llnl.gov}

\author[0000-0002-3259-2771]{Leigh C. Smith}
\affiliation{Institute of Astronomy, University of Cambridge, Madingley Rd, Cambridge CB3 0HA, UK}
\email{lsmith@ast.cam.ac.uk}

\author[0009-0004-0128-4217]{Victor Karkour}
\osu 
\email{karkour.2@buckeyemail.osu.edu}


\collaboration{20}{(Leading Authors)}

\author[0000-0002-0287-3783]{Natasha S. Abrams}
\berkeley
\email{nsabrams@berkeley.edu}







\author[0000-0001-9879-9313]{Tabetha S. Boyajian}
\lsu
\email{boyajian@lsu.edu}







\author[0000-0002-3853-7327]{Rachel B. Fernandes}
\altaffiliation{Center for Exoplanets and Habitable Worlds (CEHW) Fellow}
\affiliation{Department of Astronomy \& Astrophysics, 525 Davey Laboratory, 251 Pollock Road, Penn State, University Park, PA, 16802, USA}
\affiliation{Center for Exoplanets and Habitable Worlds, 525 Davey Laboratory, 251 Pollock Road, Penn State, University Park, PA, 16802, USA}
\email{rbf5378@psu.edu}

\author[0000-0003-0395-9869]{B. Scott Gaudi}
\osu
\email{gaudi.1@osu.edu}




\author[0000-0002-1743-4468]{Eamonn Kerins}
\manchester
\email{Eamonn.Kerins@manchester.ac.uk }



\author[0000-0002-6406-1924]{Casey Y. Lam}
\carnegie
\email{clam@carnegiescience.edu}

\author[0000-0001-9611-0009]{Jessica R. Lu}
\berkeley
\email{jlu.astro@berkeley.edu}


\author [0000-0002-1817-0329]{Carissma McGee}
\affiliation{Department of Aeronautics \& Astronautics, Massachusetts Institute of Technology, Cambridge, MA 02139, USA}
\email{cmcgee32@mit.edu}


\author[0000-0002-7669-1069]{Sebastiano Calchi Novati}
\ipac
\email{snovati@ipac.caltech.edu}



\author[0000-0002-3481-9052]{Keivan G.\ Stassun}
\vanderbilt
\email{keivan.stassun@vanderbilt.edu}




\author[0000-0002-5029-3257]{Sean K. Terry}
\umd
\email{skterry@umd.edu}

\author[0000-0002-3612-5628]{Emelly D. Tiburcio}
\lsu
\email{etibur1@lsu.edu}

\author[0000-0002-6302-251X]{Himanshu Verma}
\lsu
\email{hverma@lsu.edu}


\author[0000-0003-2872-9883]{Farzaneh Zohrabi}
\lsu
\email{fzohra2@lsu.edu}

\collaboration{51}{(Roman Galactic Exoplanet Survey Project Infrastructure Team)}

\begin{abstract}

The optimization and interpretation of microlensing surveys depends on having an accurate model of the Milky Way. However, existing population synthesis Galactic modeling tools often perform poorly in replicating the stellar contents of the inner Galactic bulge region and reproducing microlensing survey results. We present an updated Galactic model implementation within the \synthpop framework that has been tuned for simulating the upcoming {\it Nancy Grace Roman Space Telescope}'s Galactic Bulge Time Domain Survey (RGBTDS). We evaluate the model against stellar catalogs and kinematics from optical and infrared surveys toward the Galactic bulge, finding good agreement in much of the bulge, including the RGBTDS' contiguous lower bulge fields. However, within Galactic latitudes of $b\lesssim0.5^\circ$ of the Galactic plane, some inconsistencies arise which may impact projections for the RGBTDS' Galactic center field. 
The model over-predicts optical microlensing event rate per star measurements by a $\sim20$\%, but detailed comparisons to near-infrared measurements are hampered by their lack of detection efficiencies. 
{\it Roman}'s GBTDS and Galactic Plane Survey will be instrumental in resolving the remaining model inconsistencies and improving our understanding of the structure of the central few degrees of our Galaxy.

\end{abstract}


\section{Introduction} \label{sec:intro}

The \textit{Nancy Grace Roman Space Telescope} is set to launch in September 2026, with one of its three key surveys being a deep, high-cadence survey toward the Galactic bulge, the Galactic Bulge Time Domain Survey (GBTDS) \citep{rotac_report}. 
The primary science case of the \iroman GBTDS is to use gravitational microlensing to characterize the demographics of cold and free-floating exoplanets \citep[see e.g.][]{Penny2019,Johnson2020}. Gravitational microlensing occurs when a massive ``lens'' object passes in close alignment in front of a luminous ``source'' object, bending its light. This causes a transient photometric brightening and a (typically undetectable, $\lesssim$ mas) astrometric shifting signal. 
The microlensing technique is uniquely sensitive to cool planets and dim host stars because it relies only on their mass, not luminosity.  
From a photometric microlensing signal alone, the only routine observable is the timescale, which is a degenerate combination of the lens mass, distance, and proper motion. This degeneracy may be broken via astrometric microlensing signal detection, detections of the lens flux, microlensing parallax, finite source effects, and/or by using Galactic model-based priors. Additionally, Galactic models have been used to optimize the survey \citep[e.g.][]{gbtds-ccsdc-report} and validate its exoplanet mass measurement requirements \citep[e.g.][]{Terry2026}.

\citet{Kerins2009} first illustrated the utility of population synthesis Galactic models in producing maps of microlensing event rates and optical depths across the sky, using the Besan{\c c}on model \citep{Robin2003}. \citet{Penny2019} used version 1106 of the Besan{\c c}on Galactic model, an intermediate version between those published by \citet{Robin2003} and \citet{Robin2012b}, to estimate the number of microlensing events and planet detections expected for the \iroman GBTDS. 
\citet{Penny2019} and \citet{Terry2020} discussed shortcomings of current Galactic models used for microlensing simulations, noting the need for improvement to accurately optimize and interpret the RGBTDS. 

\citet{Koshimoto2021model} made significant progress on this front, presenting a new Galactic model optimized for microlensing simulations based on data from several kinematic studies and photometry and microlensing rates from the Optical Gravitational Microlensing Experiment \citep[OGLE, phases III \& IV;][]{Udalski2003, Udalski2015}. \citeauthor{Koshimoto2021model} paid particular attention to the kinematics of the disk and the IMF of the bulge, which are often neglected by other models. The model code is available publicly as {\tt genstars}\footnote{\url{https://github.com/nkoshimoto/genstars}} \citep{Koshimoto2022}.

In parallel, the \synthpop (\citetalias{synthpopI}~\citeyear{synthpopI}) Galactic modeling framework was constructed as a more flexible alternative to existing codes, where components could be easily modified and improved over time, for \iroman preparation and analysis. The model components from \citet{Koshimoto2021model} have since been made available within the \synthpop framework.
We present our model built in \synthpop from modified and combined components of existing Galactic models in \S\ref{sec:model}. \S\ref{sec:tests}-\ref{sec:mulens} present evaluations of the model against observational data. We discuss the model's overall performance and directions for future improvement in \S\ref{sec:disc} and conclude in \S\ref{sec:concl}. This is Paper I in a series of two, where Paper II will apply this model to make updated predictions for stellar and compact object microlensing events in the upcoming \iroman GBTDS. 

\section{Model Description} \label{sec:model}
We use the \synthpopp\footnote{\href{https://github.com/synthpop-galaxy/synthpop}{https://github.com/synthpop-galaxy/synthpop}} (\citetalias{synthpopI}~\citeyear{synthpopI}) Galactic population synthesis modelling code release version v1.1.3 \citep{huston_2026_20334473} to generate synthetic star catalogs based on a customized Galactic model. 
We refer to the model defined here as SP-H25, indicating the model being constructed in SynthPop, led by first author Huston, and finalized in 2025. The coordinate system is defined by default SynthPop's Solar position and motion parameters (see \citetalias{synthpopI}~\citeyear{synthpopI}, Table 1 \& Appendix A).

Prior literature examining microlensing distributions with a Galactic model have predominantly used the Besan{\c c}on model \citep{Robin2003,Robin2012b}. When compared to ground-based microlensing surveys, the model version 1106 used in previous {\it Roman} and {\it Euclid} microlensing simulations \citep{Penny2013,Penny2019} tends to under-predict event rates per unit area by a factor of $\sim2-3$, though more recent event rate measurements would place it closer to the lower end of this. In our model development, we applied some of the Besan{\c c}on model's foundation and pulled in additional recent developments in our understanding of the Milky Way's structure to make more accurate predictions of microlensing survey yields. The final model is available for use in \synthpop now, with the ``Huston2025'' model and the ``huston2025\_defaults.synthpop\_conf'' default configuration file. Its components are summarized in Table \ref{tab:model}.

\setlength{\extrarowheight}{0pt}
\begin{deluxetable}{p{2cm}p{7.5cm}p{7cm}}
    \tablehead{\colhead{Model component} & \colhead{Description} & \colhead{Reference}}
    \tablecaption{SP-H25 \synthpop model summary. Each stellar population has its own mass density distribution, kinematics, age, and metallicity distribution. The isochrones, initial mass function, initial-final mass relation, and extinction are applied universally.    \label{tab:model}}
    \startdata
    Bulge & Density: triaxial bulge (E3)  & \citet{Cao2013,Dwek1995} \\
      ~ & Kinematics: solid body bar rotation, streaming along the bar, \& position-dependent velocity dispersions & \citet{Koshimoto2021model} \\
      ~ & Age: 10 Gyr & \citet{Robin2003} \\
      ~ & Metallicity: Double Gaussian [Fe/H]: -0.31$\pm$0.31 \& 0.26$\pm$0.2 dex & \citet{Gonzalez2015} \\ 
      \tableline
    Nuclear Stellar Disk & Density \& kinematics: quasi-isothermal distribution function & \citet{Sormani2022} \\
      ~ & Age \& Metallicity: same as bulge & ~ \\
      \tableline
    Thin disk & Density: sech$^2$-scaled disk & \citet{Koshimoto2021model} \\
      ~ & Kinematics: modified \citet{Shu1969} distribution function & \citet{Koshimoto2021model, Sharma2014,Bland-Hawthorn2016} \\
      ~ & Age: 7 subpopulations between 0-10 Gyr & \citet{Robin2003} \\
      ~ & Metallicity: Gaussians with [Fe/H] means from -0.37$-$0.03 and dispersions from 0.10$-$0.20 & '' \\
      \tableline
    Thick disk &  Density: exponential disk & \citet{Koshimoto2021model} \\
      ~ & Kinematics: same as thin disk & '' \\
      ~ & Age: 12 Gyr & '' \\
      ~ & Metallicity: Gaussian [Fe/H] = -0.78$\pm$0.30 & \citet{Robin2003} \\
      \tableline
    Halo & Density: spheroid & \citet{Robin2003} \\
      ~ & Kinematics: zero-mean Gaussians & '' \\
      ~ & Age: 14 Gyr & '' \\
      ~ & Metallicity: Gaussian [Fe/H] = -1.87$\pm$0.50 & '' \\
    \tableline \\[-4ex]
    \tableline
    Isochrones & MIST v1.2 with rotation v/vcrit=0.4 & \citet{Dotter2016,Choi2016} \\
    Initial mass function & $\xi(m)\propto m^{-\alpha}$, with $\alpha=1.3$ for $0.08\leq m/M_\odot<0.5$, $\alpha=2.3$ for $m/M_\odot\geq0.5$  & \citet{Kroupa2001} \\
    \multirow{2}{2cm}{Initial-final mass relation} & White dwarfs: $m_{\rm WD} = 0.109m_{\rm init} + 0.394M_\odot$ & \citet{Kalirai2008} \\
     ~ & Neutron stars: Gaussian $m_{\rm NS} = 1.36 \pm 0.09 M_\odot$ & \citet{Rose2022} \\
     ~ & Black holes: SukhboldN20 metallicity-dependent, 2-branch scheme from PopSyCLE & \citet{Rose2022,Sukhbold2014,Sukhbold2016,Woosley2017,Woosley2020} \\
    Extinction & Map: 2-d map at the Galactic bulge, scaled along each line of sight as an exponential disk & \citet{Surot2020,Sharma2011} \\
     ~ & Law: updated \citet{Cardelli1989} law, the ``SODC'' SynthPop module with R$_V$=2.5 & \citetalias{synthpopI} (\citeyear{synthpopI}); \citet{O'Donnell1994,Surot2020}
    \enddata
\end{deluxetable}

\subsection{Stellar Densities \& Kinematics}

The SP-H25 model's stellar mass distribution functions pull from existing models and observational work. \citet{Cao2013} analyzed red clump giants in the OGLE-III database \citep{Nataf2013} to construct a Milky Way Galactic bar/bulge model, comparing and fitting selected triaxial bulge models from \citet{Dwek1995}. Their preferred parameterization is the E$_3$, where stellar density is modeled as a Bessel Function of the second kind with independent scale lengths along x-, y-, and z-axes (0.67, 0.29, and 0.27 kpc, respectively) and including the angle of the bar from our line of sight as a fit parameter (29.4$^\circ$). 
The \citet{Cao2013} bulge model does not include kinematics, so we adopt the bulge kinematic model from the \cite{Koshimoto2021model}, which applies a Gaussian distribution with a mean velocity and velocity dispersion that vary with Galactic position. The mean velocity is characterized by solid-body rotation of the bar plus streaming motion along the bar. The kinematic model parameters for a bulge defined by a Bessel function of the second kind were provided by N. Koshimoto (private communication). As in the \citet{Koshimoto2021model} model, the bulge model is not dynamically self-consistent, but the velocity dispersion peaking in the Galactic Center and gradually decreasing with distance is consistent with dynamical models, as shown by \citet{Sanders2019}.

In addition to the bulge kinematic model, we adopt the \citet{Koshimoto2021model} disk model both for stellar density and kinematics. The disk is divided into 7 age-based sub-components of the thin disk and one additional thick disk component, following the disk sub-component scheme used in the Besan{\c c}on model \citep{Robin2003}. The disk models are flat in the inner Galaxy, where the disk is less well-constrained, and a sech$^2$ (thin disk) or exponential (thick disk) function in the outer (R$\gtrsim$5 kpc) disk. 
Compared against Galactic models used in prior microlensing analysis, \citeauthor{Koshimoto2021model} included a more detailed model of disk kinematics motivated by Gaia DR2 \citep{GaiaCollaboration2018}.
The disk kinematic model parameterization is drawn from \citet{Sharma2014} and includes a skewed azimuthal velocity distribution. It assumes that the Galaxy is in dynamical equilibrium and a uses a modified Shu distribution function \citep{Shu1969, Schonrich2012, Sharma2013}. In short, it probabilistically selects a guiding-center radius for each star as a function of its true Galactocentric distance. Rotation velocity is drawn from a rotation curve based on \citet{Portail2015}, and azimuthal velocity from a function of rotation velocity, guiding-center radius, and true Galactocentric distance. Velocity dispersions in the $R$ and $z$ dimensions are computed as a function of Galactocentric distance and population age.

The stellar halo is unimportant for microlensing survey statistics in the direction of the Galactic bulge and plane ($\sim0.1$\% of all stars in an example \iroman GBTDS field). We adopt the halo spheroid density and zero-mean Gaussian distribution kinematic models of the Besan{\c c}on model \citep{Robin2003}. As with the other populations in our model, compact objects in the stellar halo are generated as a part of this population, where stars that have evolved beyond the phases covered by the isochrones (see \S\ref{sec:model_evol}) are kept as dark compact objects (white dwarfs, neutron stars, and black holes with no photometry). In the halo, this is $\sim15$\% of the population's generated stars. No additional compact halo objects are injected.

\citet{Navarro2020} analyzed the Galactic latitude dependence of microlensing event rate per unit area on-sky from the VVV survey, noting that the completeness-corrected event count in the infrared (K-band) peaks in the Galactic plane, which motivates targeting the Galactic center with \iromann. \citet{Wen2023} noted the same latitude trend in observed raw (not corrected for detection efficiency) H- and K-band event rate per star from the UKIRT survey. Stellar density is the highest in the Galactic plane and center, but dense dust extinction in the plane causes a dip in microlensing events and star counts at optical wavelengths. 
\citet{Kerins2009} demonstrated the diminishing impact of extinction on event rates with increasing wavelength in simulations (I- to J- to K- band), but, contrary to the VVV and UKIRT observations, their model still predicted a $\sim$50\% dip in the plane even for K-band event rates per square degree. 
With disagreements between existing models and data, it is not clear at what wavelength the effect of extinction diminishes enough that this dip disappears. With the GBTDS' primary filter being a wide-YJH filter in between the optical and K-band wavelengths, significant uncertainties remain in the expected event rates near the Galactic Center.

In addition to the effects of extinction, other populations besides the halo, bulge, and disk exist near the Galactic center and contribute to the underlying star counts. \citet{Kondo2023} found that {\tt genstars}-simulated H-band event rates per source star peak near the plane when the Nuclear Stellar Disk (NSD) is included.

For fields near the center of the Galaxy, our model also includes this NSD model to better account for the full stellar contents of the inner latitudinal degree of the Galactic plane. 
Based on the \citet{Fritz2021} NSD survey and additional kinematic data, \citet{Sormani2022} fit an axisymmetric, self-consistent equilibrium dynamical model of the population with a total mass of $10.5\times10^8$ M$_\odot$.  In \synthpopp, the density and kinematics (average rotational velocity and velocity dispersions) for this component are interpolated across a grid tabulated via {\sc AGAMA} \citep[All-purpose GAlaxy Modelling Architecture\footnote{\url{https://github.com/GalacticDynamics-Oxford/Agama}};][]{Vasiliev2019} by N. Koshimoto (private communication) using the \citet{Sormani2022} model.

We note that some of the mass distributions adopted in our model draw from sources that model an independent white dwarf component and exclude neutron stars and black holes. In contrast, in \synthpop all stars and stellar remnants are evolved together and included as a single population, so the remnant mass density is added into the total stellar density in these cases.

\subsection{Stellar Properties and Evolution} \label{sec:model_evol}

Each population in our model (the bulge, 7 thin disk sub-components, thin disk, halo, and nuclear stellar disk) is assigned age and metallicity distributions.
The ages of stars in the bulge and halo are set to 10 and 14 Gyr, respectively, based on \citet{Robin2003}. 
We assume the NSD to be the same age as the bulge, which is consistent with the findings of \citet{Nogueras-Lara2020} that $>$80\% of NSD stars are between 8-13.5 Gyr in age. Each thin disk component draws ages from a uniform distribution across a range, adopting the values used by both \citet{Robin2003} and \citet{Koshimoto2021model}: 0-0.15, 0.15-1, 1-2, 2-3, 3-5, 5-7, and 7-10 Gyr. Thick disk stars are all assigned an age of 12 Gyr, based on \citet{Koshimoto2021model}, which differs slightly from the 11 Gyr-old thick disk of \citet{Robin2003}. 

We use [Fe/H] to define metallicity with solar-scaled abundances and adopt the disk and halo Gaussian metallicity distributions of \cite{Robin2003}. We use an updated bulge metallicity distribution from \citet{Gonzalez2015}, parameterized as a double Gaussian function, and also apply it to the NSD. 
Many Galactic models including those from \citet{Robin2003} and \citet{Koshimoto2021model} apply piecewise power law initial mass functions (IMFs) with slight variations. We adopt the IMF of \cite{Kroupa2001} and allow for the generation of stars from 0.08-100M$\odot$, limiting to stellar masses with brown dwarfs to be added in future model developments. 

Stellar evolution and photometry are estimated from the MIST (MESA\footnote{Modules for Experiments in Stellar Astrophysics \citep{mesa1,mesa2,mesa3,mesa4,mesa5,mesa6}} Isochrones and Stellar Tracks; \citealp{Choi2016,Dotter2016}) version 1.2 isochrones, in order to maximize the metallicity and age parameter space covered, as well as photometric filters available. We use the v/vcrit=0.4 packaged model grids for detailed stellar properties from their theoretical and synthetic photometry isochrones. We use the {\sc CharonInterpolator} described by \citetalias{synthpopI}~(\citeyear{synthpopI}) to smooth interpolation in the red giant phase. 
For stars between 0.08-0.1M$_\odot$, which lie outside of the MIST grids, we draw initial masses, ages, and metallicities, then assume zero mass loss and do not generate photometry, allowing these stars to be dark lenses but not sources in our microlensing simulations. 

The MIST isochrones trace stellar evolution through either the carbon burning phase or into a white dwarf cooling sequence, depending on initial mass. For stars with ages beyond those covered in their respective tracks, we apply an initial-final mass relation (IFMR) to assign realistic masses to dark stellar remnants including post-cooling track white dwarfs (WDs), neutron stars (NSs), and black holes (BHs). They are not assigned photometry and are considered only as lenses in a microlensing simulation.
For WDs, we adopt the \citet{Kalirai2008} IFMR:
$m_{\rm WD} = 0.109m_{\rm init} + 0.394M_\odot$,
where the WD remnant type for stars with initial mass $m_{\rm init}<9M_\odot$.

For high-mass ($m_{\rm init}>9M_\odot$) stars that have evolved beyond the MIST grids, we adopt the SukhboldN20 IFMR from {\sc PopSyCLE} \citep[][\S3.4]{Rose2022}. It is based on zero-metallicity models from \citet{Sukhbold2014}, solar-metallicity models from \citet{Sukhbold2016}, and pulsational-pair instability models from \citet{Woosley2017, Woosley2020}. While the high-mass IFMR is not observationally well-constrained, this method provides a theoretical prediction from detailed simulations. 
The compact remnant type (NS or BH) is assigned probabilistically according to mass bins. 
NS masses are drawn from a Gaussian distribution with mean 1.36$M_\odot$ and standard deviation 0.09$M_\odot$. 
BH masses come from a piecewise function that splits at $m_{\rm init} = 39.6M_\odot$. The higher-mass branch of the BH IFMR is metallicity-dependent, with higher metallicity resulting in lower $m_{\rm final}$. The lower-mass branch of the IFMR has no metallicity dependence. For any object where the BH track results in $m_{\rm final}<3M_\odot$, the remnant is reassigned to be a NS and given a new mass from the NS distribution.

\subsection{Extinction}

Extinction is critical for accurately modeling microlensing event distributions in regions near the Galactic bulge and plane where the dust content is significant. Extinction maps may be provided in 2- or 3-dimensions. Three-dimensional maps are well-suited for incorporation into a population synthesis Galactic model, providing extinction in a reference filter as a function of angular position on-sky and distance; however, they often suffer from diminishing accuracy at large distances, particularly in regions of high extinction and/or low stellar densities \citep[see e.g.][]{Zucker2025}. Two-dimensional maps, on the other hand, may provide accurate values for objects beyond the Milky Way, or in the Galactic bulge, depending on analysis method; however, some consideration is required in how to apply these to 3-dimensional distributions of Milky Way stars. Early work on \synthpop primarily used the 3-d dust map from \citet{Marshall2006}, but due to the limitations mentioned above, it often fails to reproduce observed values in optical color-magnitude space toward the Galactic bulge. 


The online Besan{\c c}on model generator\footnote{\url{https://model.obs-besancon.fr/}} allows for a few different extinction implementations: a diffuse thin disc (not reliable near the Galactic plane), the \citet{Marshall2006} 3-d map, or a user-input line of sight distribution. 
The {\sc Galaxia} code, which is a modified implementation of the Besan{\c c}on model, uses the 2-d extinction map from \citet{Schlegel1998} as the extinction value at infinity, and scales extinction with distance assuming that dust is distributed as a double-exponential disk \citep[][see section 2.6]{Sharma2011}.
The updated {\tt genstars} code described initially by \citet{Koshimoto2021model} applies the 2-dimensional maps from \citet{Gonzalez2012} and \citet{Surot2020} to calculate extinction at the mean Galactic bulge red clump star distance and, similarly to Galaxia, scales foreground and background extinction from this value according to an exponential disk dust distribution. 

We adopt a similar scheme to the latter two, using the 2-dimensional extinction map from \citet{Surot2020} as the true extinction values at 8.15 kpc (our assumed Galactic center distance), transformed into 3-d via the {\sc Galaxia} dust disk scaling. The \citet{Surot2020} map is limited to $-10^\circ<l<10^\circ$, $-10^\circ<b<5^\circ$. This area contains most of the Galactic bulge, and importantly, the \iroman Galactic Bulge Time Domain Survey region. Thus, we limit our analysis in this work to this region of sky. Users may adapt this model to other regions of sky by substituting a different extinction map from the many options in \synthpop that better is optimized for their purposes.

We adopt the SODC \synthpop extinction law, which adopts the framework of \citet{Cardelli1989}, with an optical law based on \citet{O'Donnell1994} and infrared law adjusted to be consistent with the \citet{Surot2020}, \S2.5) reddening vector (see also \citetalias{synthpopI}~\citeyear{synthpopI}. We adopt the total-to-selective extinction ratio of $R_V=2.5$, as measured via infrared and optical data toward the bulge by \citet{Nataf2013}.

\subsection{Preliminary SP-H24 Model}
A prior version of this model was used in several past and ongoing Roman microlensing modelling efforts \citep{gbtds-ccsdc-report, rotac_report, Terry2026, Zohrabi2026}. We refer to this model as SP-H24, indicating the model being constructed in \synthpopp, led by first author Huston, and finalized in 2024.
The model had two key differences from the current version: the exclusion of a nuclear stellar disk component, and a simplified extinction model. The SP-H24 extinction model applied the 2-d \citet{Surot2020} map as a screen at 2 kpc. Additionally, \synthpop code version (v1.0.0) used for the SP-H24 simulations contained a bug in the kinematic module for bulge stars, which primarily impacted radial velocities but also had a small impact on $\mu_l$ proper motion, worsening at larger $|l|$ with minimal impact in the \iroman GBTDS fields.

\section{Model Evaluation Approach}\label{sec:tests}

\begin{figure}
    \centering
    \includegraphics[width=\linewidth]{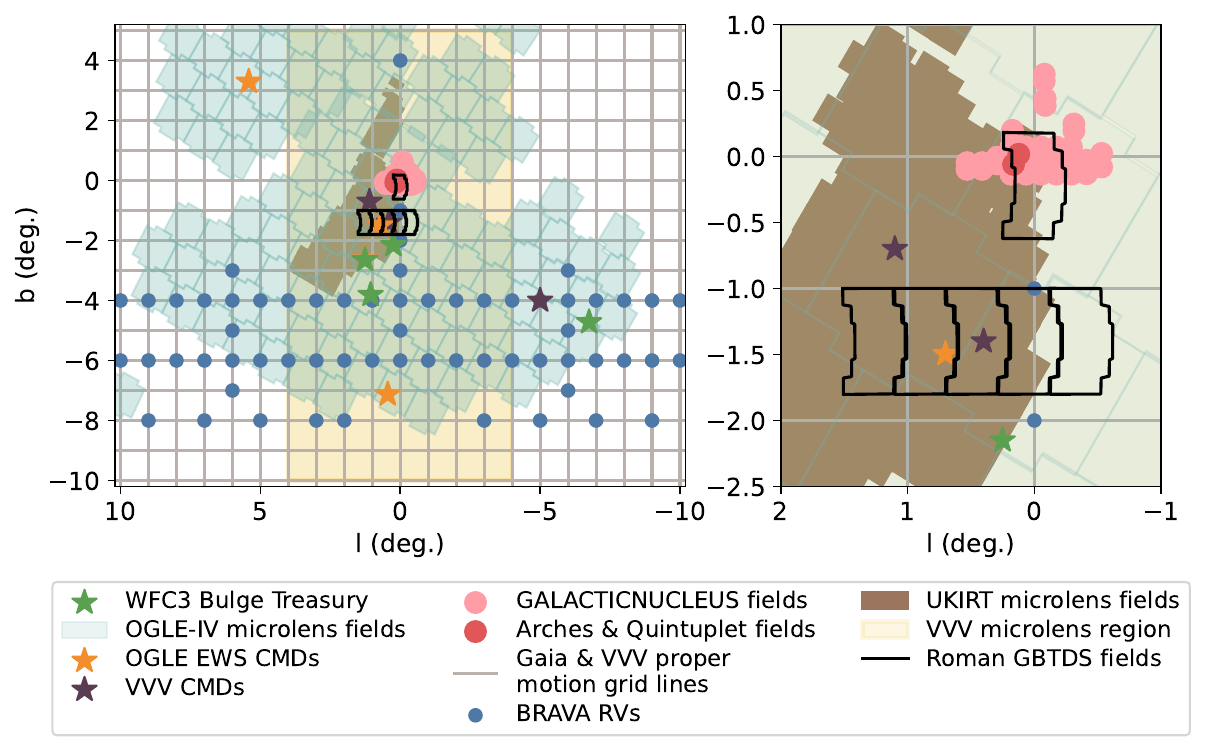}
    \caption{Locations on-sky of the observational data sets used to evaluate the SP-H25 model across the full region covered by our extinction map (left) and zoomed in on the \iroman GBTDS fields (right). Stars mark scattered bulge fields used for luminosity function of color-magnitude diagram comparisons, while circles mark those focused on the Galactic Center region. The 1$^\circ$ grid lines mark the proper motion sampling for wide-area surveys, and triangles mark the radial velocity sample fields. Past microlensing survey fields are shaded in, and the \iroman GBTDS fields are outlined in black.}
    \label{fig:test_fields}
\end{figure}

We compare catalogs generated with the SP-H25 model to various data sets for validation. In $\S$\ref{sec:counts}, we examine optical and infrared star counts, luminosity functions, and color-magnitude distributions to validate primarily the stellar density distributions and the extinction model. In $\S$\ref{sec:kine}, we examine proper motion and radial velocity distributions to test primarily the kinematic models and also discuss how other model aspects affect these observables. In $\S$\ref{sec:mulens}, we compare simulated microlensing event distributions to microlensing survey results. Because microlensing observables are a complex function of stellar distributions, properties, kinematics, and extinction, it is difficult to draw any conclusions on individual model aspects from these evaluations. However, evaluating the model's ability to reproduce existing microlensing results is valuable in determining how reliable the model may be for \iroman microlensing survey simulations.
The on-sky positions of the observations used in this work are shown in comparison to the GBTDS fields in Figure \ref{fig:test_fields}. 

We prioritize comparisons to completeness-corrected data sets, but where these are not available, we examine uncorrected catalog-level data as well. We compare completeness-corrected observational data sets directly to SynthPop output catalogs. For ground-based, uncorrected catalogs, we approximate some observational effects by artificially blending SynthPop catalog stars based on a blend radius (half of the typical seeing FWHM at the observing site), iterating over each star from brightest to dimmest to find its neighbors within the radius and absorb them into a single blended source. We sum the magnitudes for each blended set and adopt flux-weighted mean proper motions. This is not equivalent to an artificial star-based completeness-correction but is our method to mitigate easily replicable effects from the simulation side. For Gaia and HST, we ignore blending given the high spatial resolution of their space-based observations.

\section{Luminosity, Color, and Star Count Evaluations} \label{sec:counts}

\subsection{The WFC3 Galactic Bulge Treasury Program} \label{sec:bulge_treasury}
The WFC3 Galactic Bulge Treasury Program \citep{Brown2009, Brown2010} imaged four fields throughout the Galactic bulge with the Hubble Space Telescope's Wide Field Camera 3 (WFC3) in 5 optical and near-infrared filters: F390W (Washington C), F555W (V), F814W (I), F110W (J), and F160W (H). The observations were performed in four fields across the Galactic bulge region, probing the bulge and disk populations at varying levels. To compute observed luminosity functions (LFs), we fetched the v2.0 photometry catalogs\footnote{\url{https://archive.stsci.edu/prepds/wfc3bulge/}} and used the supplementary artificial star catalogs to estimate completeness per magnitude bin for each field and filter. We sampled the artificial star catalog such that the output magnitude distribution matched the distribution of the observed star catalog magnitudes, then estimated completeness as the ratio of input to output artificial stars per bin. We assume Poisson uncertainties.

\begin{figure}
    \centering
    \includegraphics[width=\linewidth]{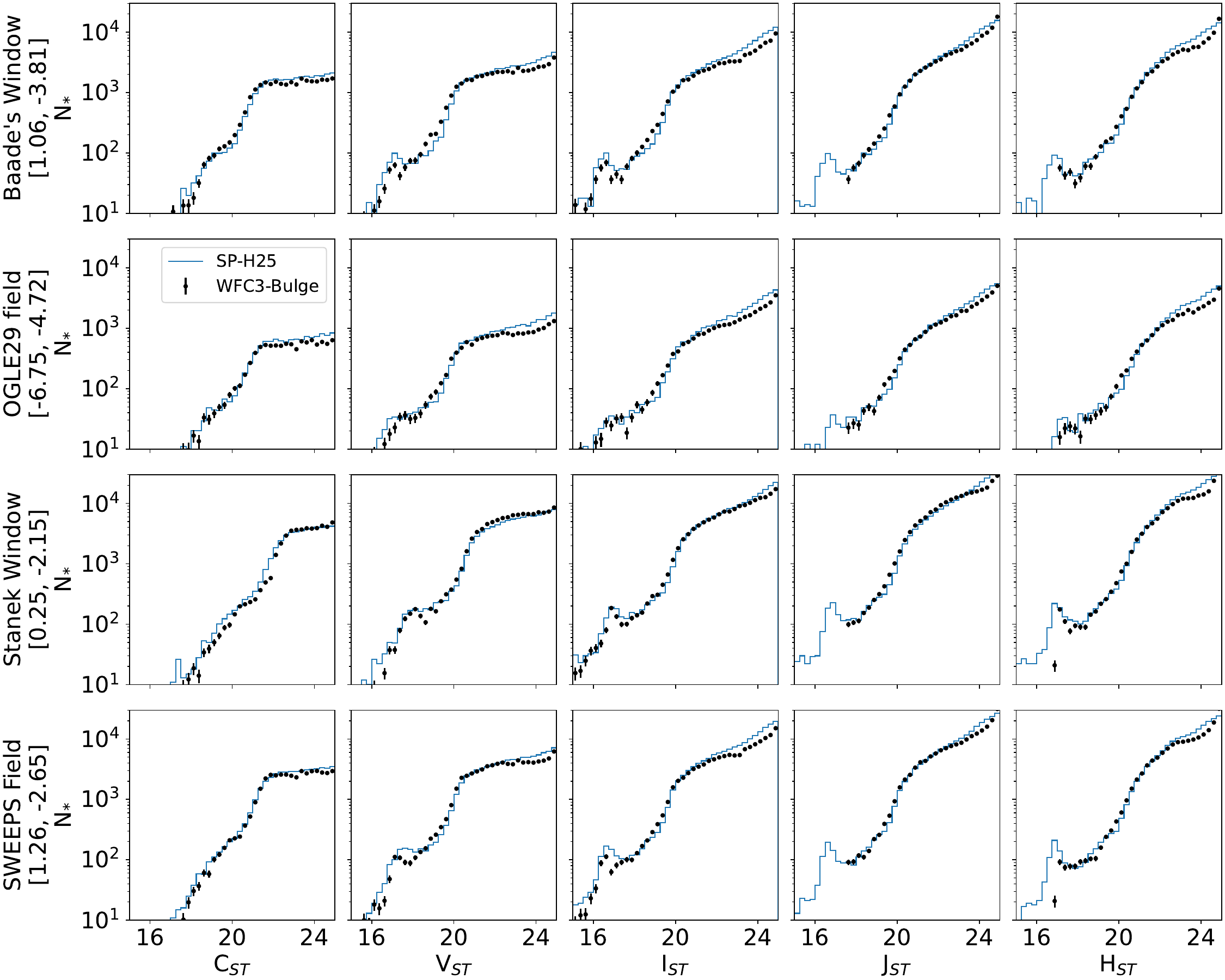}
    \caption{Observed and simulated luminosity functions for the WFC3 Galactic Bulge Treasury Program \citep{Brown2009,Brown2010} fields in STMAG \citep{Koornneef1986}.}
    \label{fig:treasury_comparison}
\end{figure}

In Figure \ref{fig:treasury_comparison}, we show SP-H25 output catalogs in comparison to these corrected observed LFs.
The LFs show good agreement along each sightline both in total stellar counts and in key features like the red clump position. The predicted stars per magnitude bin is within $20$\% of the observed value for most bins, filters, and fields. The functions deviate the most at the very bright and very dim ends. The bright end is subject to saturation on the observational side and small number statistics from both simulation and observation. The dim side disagreement may reflect inaccuracies in the low mass IMF and/or our model's lack of binarity. Deviations in the red clump position (the shelf between $\sim$17-20 mag.) of some V-band LFs likely indicate shortcomings in the extinction model, due to finite spatial resolution and/or the unphysical line-of-sight scaling scheme.

\citet{Terry2020} re-analyzed select WFC3 Bulge Treasury program data in the context of the RGBTDS, focusing on the Stanek Window, the field closest to the RGBTDS fields. They concluded that the Besan{\c c}on v1612 \citep{Robin2003,Robin2012b} and Galaxia v0.7.2 \citep{Sharma2011} Galactic models underpredict bulge star counts by $\sim33-75$\% in I- and H- band, while GalMod \citep{Pasetto2016,Pasetto2018,Pasetto2019} overpredicted star counts but only slightly ($\sim5-10$\%) in H-band. \citet{Kondo2023} compared their {\tt genstars} model H-band star counts to these as well and found good agreement until $H>19.5$, where their model under-predicts dim star counts.

\begin{figure}
    \centering
    \includegraphics[width=0.9\linewidth]{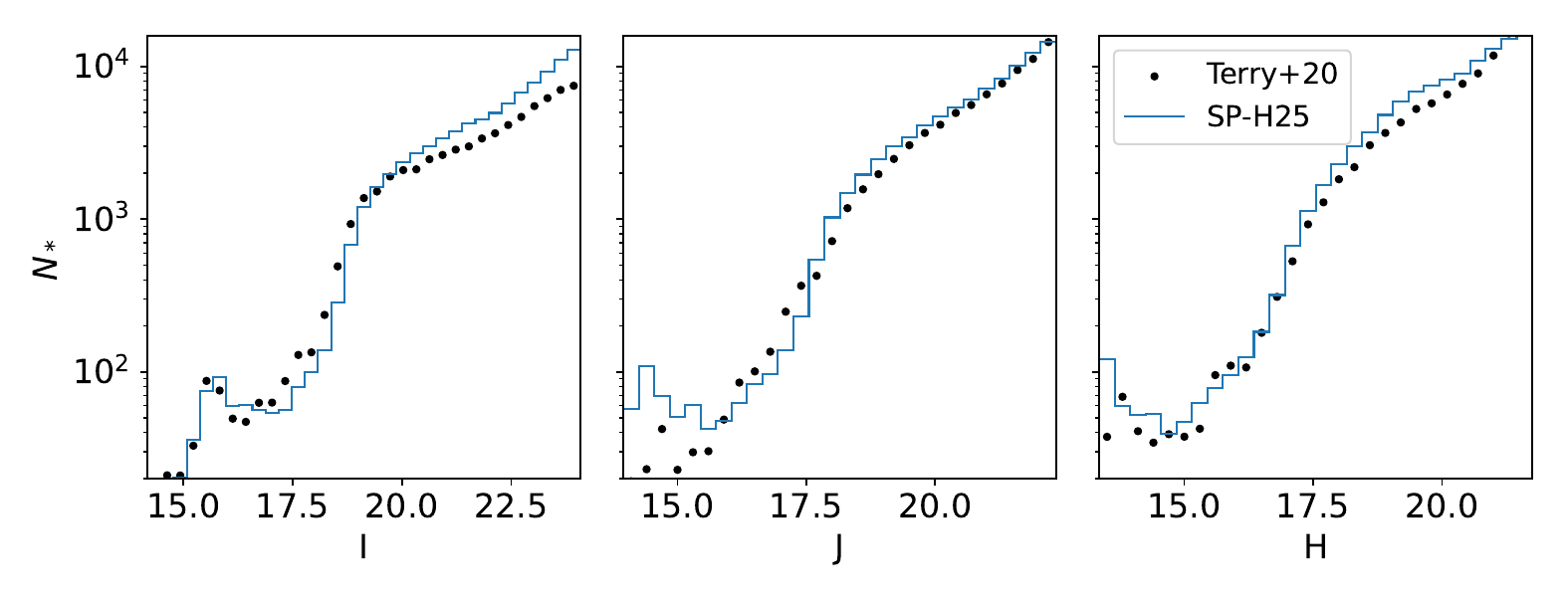}
    \caption{Observed \citep{Terry2020} and simulated bulge-only luminosity functions for the Stanek Window ($l=0.25^\circ$, $b=-2.15^\circ$). Magnitudes are shown in the VEGAMAG system.}
    \label{fig:treasury_stanek}
\end{figure}

\citet{Terry2020} trimmed their stellar catalog by proper motion to select only bulge stars and corrected for photometric completeness and proper motion cut efficiency. 
With \synthpopp, we can identify bulge stars by their population of origin in the simulation rather than a proper motion cut.
In Figure \ref{fig:treasury_stanek}, we compare selected \citet{Terry2020} Table 5 bulge-only LFs (no uncertainties provided) to a bulge-only SP-H25 catalog. Here, we see an average of 7\% over-prediction in J-band counts per bin and 22\% in H-band counts per bin.

\subsection{Optical Gravitational Lensing Experiment Star Counts} \label{sec:ogle_cmd}

\begin{figure}
    \centering
    \includegraphics[width=0.9\linewidth]{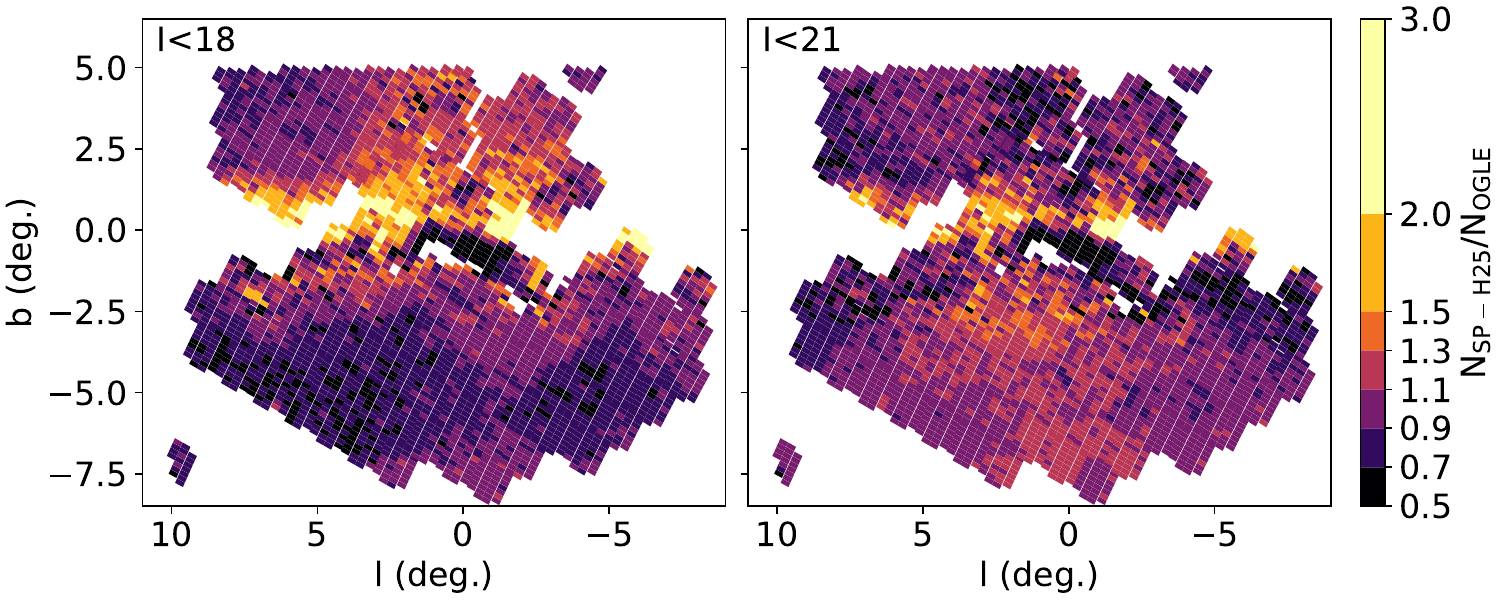}
    \caption{Star count comparison for I$<$18 (left), I$<$21 (right), comparing OGLE-IV results from \citet{Mroz2019} and SP-H25 model generated catalogs. The colormap shows the ratio of model star counts to observed star counts. }
    \label{fig:ogle_counts}
\end{figure}

\citet{Mroz2019} provided completeness-corrected star counts down to I$=$21 (Vega) across the Optical Gravitational Lensing Experiment Phase IV (OGLE-IV) bulge fields. In Figure \ref{fig:ogle_counts}, we compare these values to simulated SP-H25 catalogs for the OGLE-IV survey chips covered by our model's extinction map.
The SP-H25 model significantly overpredicts OGLE's star counts at $|b|\sim0^\circ$, but we note that these regions host the fewest optically-detectable stars and are thus subject to more significant Poisson uncertainties on the observational side and larger extinction map \& law uncertainties on the modeling side. At $|b|>1^\circ$, the star counts are generally agree within $\sim$25\%. For an OGLE field selection approximating the RGBTDS fields ($-1.5^\circ<l<0.8^\circ$, $-1.8^\circ<b<-1.0^\circ$), the SP-H25 model under-predicts I$<$18 star counts by an average of 2\% and I$<$21 by 7\%.

\subsection{Color-Magnitude Diagrams from OGLE \& VVV}

Following the approach taken by \citet[][see Appendix B]{Lam2020} to evaluate extinction laws, we compare simulated color-magnitude diagrams to those from a selection of OGLE fields. The optical observational data are provided by the OGLE Early Warning System\footnote{\url{https://ogle.astrouw.edu.pl/ogle4/ews/ews.html}} \citep[EWS][]{Udalski2015} with star catalogs for each alerted event in its respective field. We selected one field to lie in the RGBTDS lower bulge footprint and three others somewhat arbitrarily while ensuring coverage of differing levels of extinction and stellar density. Because there is no completeness correction for these observation data, we generate SP-H25 catalogs then apply our blending scheme with a blend radius of 0.65\arcsec and limit our comparisons to bright (I$\leq$18) stars.
In order to examine the effects of extinction in the infrared, we also examine CMDs from the near-infrared VISTA Variables in the Via Lactea \citep[VVV;][]{Minniti2010} survey, sampling the VIRAC2 \citep[VVV Infrared Astrometric Catalogues Version 2;][]{Smith2025} catalog in the same OGLE target fields. For VVV, we apply a blend radius of 0.36\arcsec to our SP-H25 catalog, and limit comparisons to K$\leq$17 stars.

\begin{figure}
    \centering
    \includegraphics[width=\linewidth]{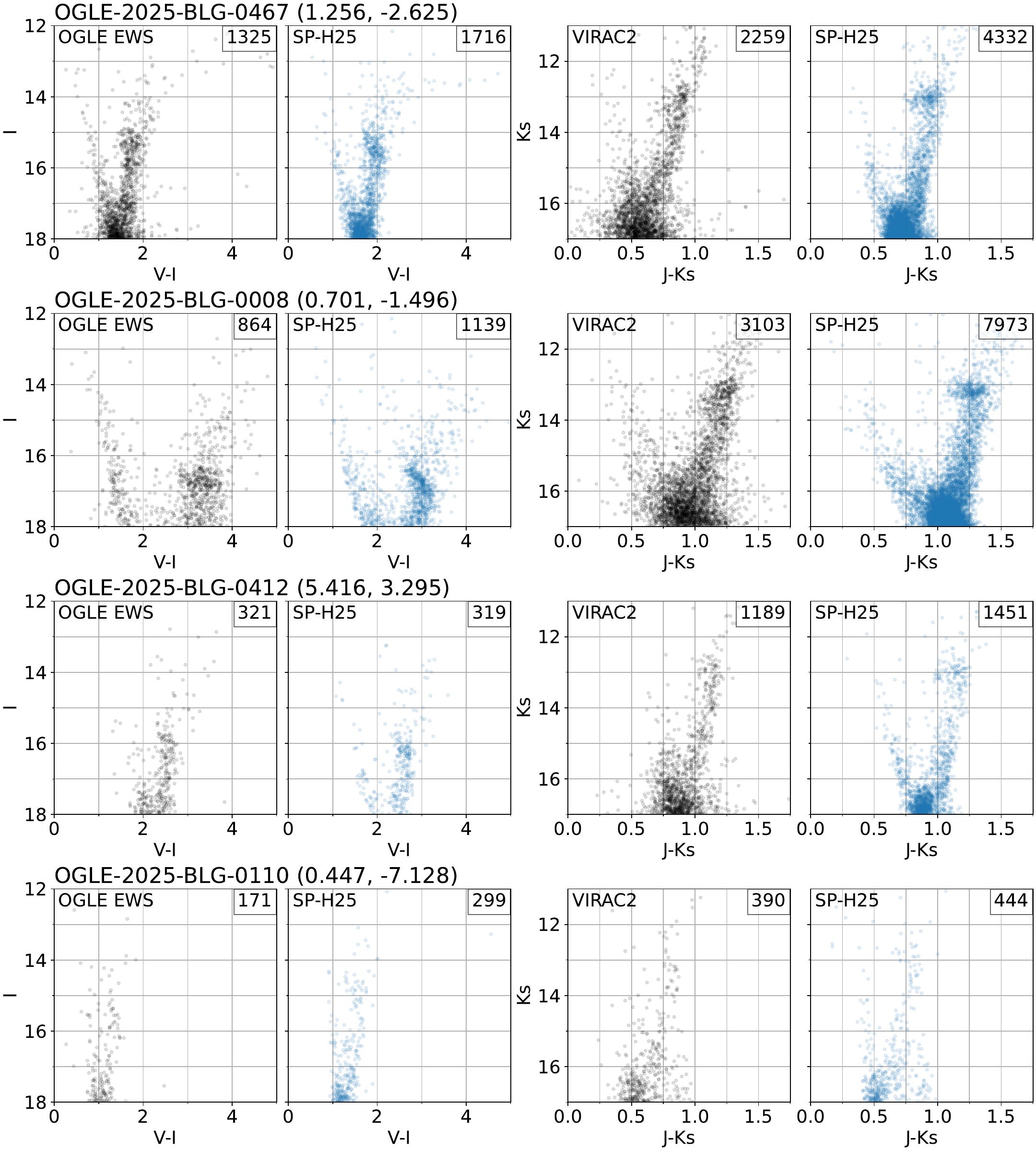}
    \caption{Comparison of color-magnitude diagrams for four sample OGLE event alert fields within the region of sky covered by our extinction map with observational data in black and SP-H25 simulations in blue. The left side shows I and V magnitudes with observational data from the OGLE EWS. The right side shows J and Ks magnitudes with observational data from the VIRAC2 catalog.}
    \label{fig:ogle_cmd}
\end{figure}

In Figure \ref{fig:ogle_cmd}, we show that the SP-H25 model replicates the key structures of these observed CMDs. The star counts are over-predicted for the most crowded fields, which to some extent is likely due to incompleteness in the OGLE EWS \& VIRAC2 catalogs. While the red clump (RC) is well-reproduced in color-space on average across all the CMDS, there is some slight variation (on the order of 1/10th mag.). Thus, the extinction map, law, and/or stellar ages and metallicities have room for future improvement. It is also visible here, particularly at dimmer magnitudes that the observational CMDs show a wider spread due to noise.

It is also notable, particularly in the top two rows of Figure \ref{fig:ogle_cmd}, that the SP-H25 model can produce a red clump (RC) that appears overly smeared in color-magnitude space when compared to observation. There are three model factors that primarily determine the red clump's shape: the spread in the [Fe/H] distribution, differential extinction (either along the line of sight or in the plane of the sky), and the stellar distances. This suggests that the model shortcomings include some combination of an overly broad metallicity distribution, an inaccurate 3-d extinction map, an underestimated bar angle, or an overestimated spatial extent of the bar. 


\subsection{Nuclear Stellar Disk}\label{sec:nsd_lfs}
\begin{figure}
    \centering
    \includegraphics[width=0.8\linewidth]{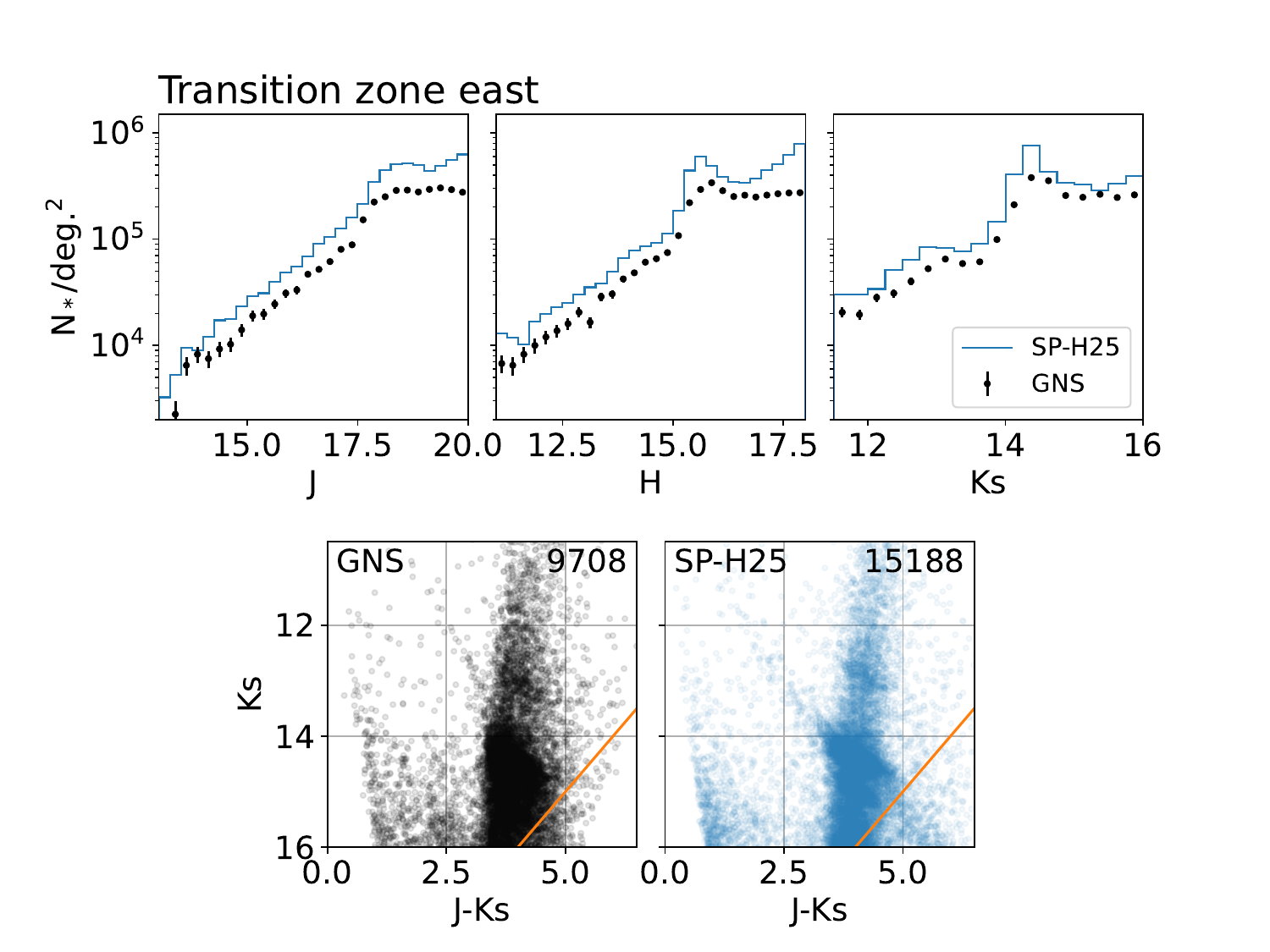}
    \caption{Near-IR luminosity functions (left 3 panels) and color-magnitude diagrams (right 2 panels) from GNS (black points) in comparison to SP-H25 model catalogs (blue). In the CMDs, an orange line marks the roughly estimated 80\% completeness limit. The ``transition zone east'' field is shown in this paper as an example, and the six additional fields will be made available in a figure set in the online version of this article.}
    \label{fig:galnuc}
\end{figure}

To test our model's performance at very low $|b|$, including the nuclear stellar disk component, we examine luminosity functions for regions closer to the Galactic center. We begin these comparisons with Data Release 1 catalogs from the GALACTICNUCLEUS near-IR survey \citep[GNS; ][]{Nogueras-Lara2019}. This survey's fields are spaced across the inner Galactic bulge and nucleus between $|l|<0.6^\circ$ and $-0.2^\circ<b\lesssim0.6^\circ$. 
Because there is no completeness correction in these observational catalogs, we generate SP-H25 star catalogs then apply our blending scheme with a blend radius of 0.1\arcsec and limit our comparisons to bright stars \citep[J$\leq$20, H$\leq$18, K$\leq$16; the roughly estimated 80\% completeness limits reported by][]{Nogueras-Lara2019}.
\synthpop v1 only generates circular fields on-sky, so we generate synthetic catalogs at the center of each GNS sub-field, and trim the GNS stars used in our comparison to those within the circular region simulated, then re-combine these catalogs into the named fields as defined by the GNS. For computational feasibility, we limit these sample fields to $3\times10^{-4}$ square degrees for the Central region and $2\times10^{-3}$ square degrees for all other regions.
Figure \ref{fig:galnuc} shows the observed LFs with Poisson uncertainties and CMDs in comparison with the simulation.

This comparison shows the most severe model-data disagreement thus far. SP-H25 overpredicts star counts by $\sim$50-100\% in many locations and filters. While the observed LFs have no accounting for completeness, it is difficult to quantify what effect this would have of the data. We adopted a blend radius based on the reported survey resolution of 0.2\arcsec~\citep{Nogueras-Lara2019}, though the image quality likely varied across fields due to observing conditions. 
Further re-analysis of the GNS observations and their completeness are underway (F. Nogueras-Lara, private communication), and a re-examination with these updates will be an important next step in \synthpopp's continued improvement. The H-band magnitude of the RC bump is consistently well-matched throughout most of the fields, while agreement is a bit less consistent at Ks-band. In the highest-extinction fields, and particularly for J-band, the RC is difficult to analyze due to the completeness drop-off at the dim ends of each LF. While the SP-H25 model over-predicts overall star counts, the color-magnitude diagrams show that it captures the key features in stellar population separation and extinction.

\begin{figure}
    \centering
    \includegraphics[width=0.8\linewidth]{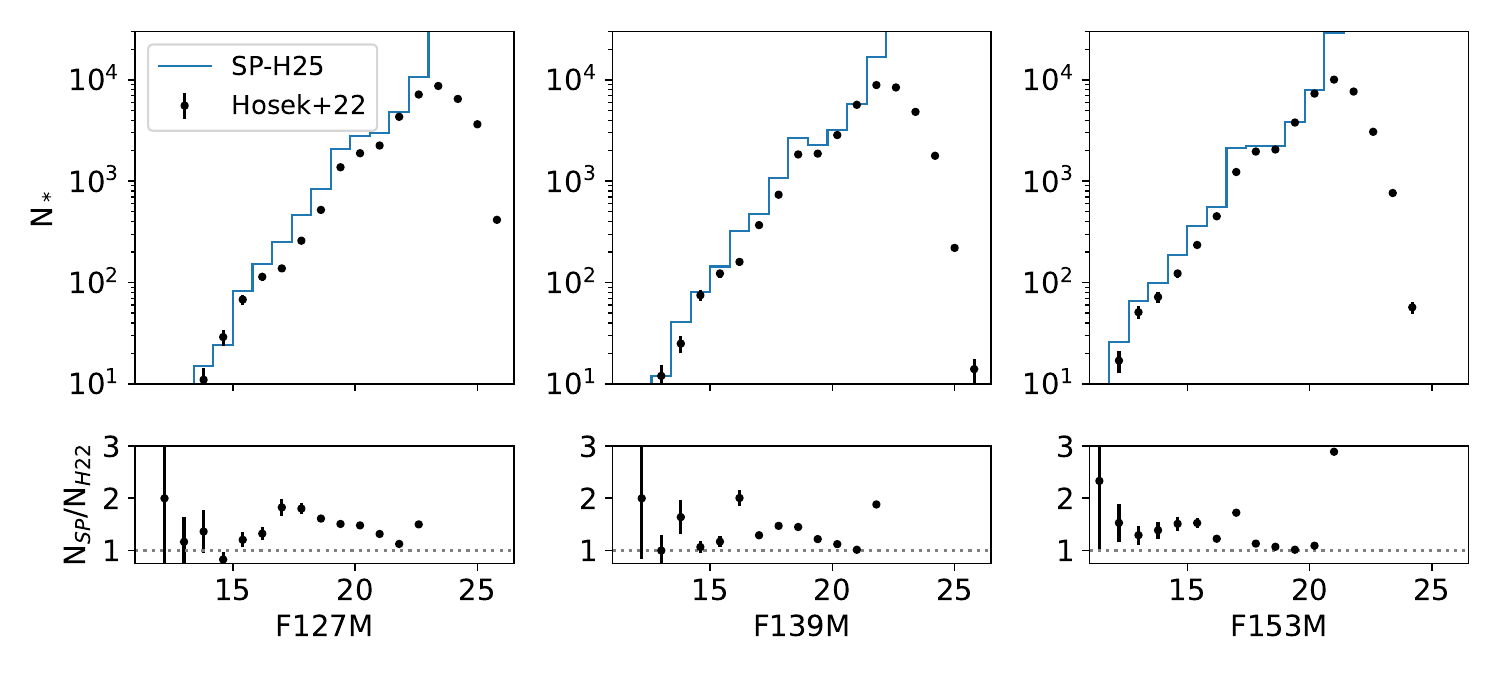}
    \includegraphics[width=0.8\linewidth]{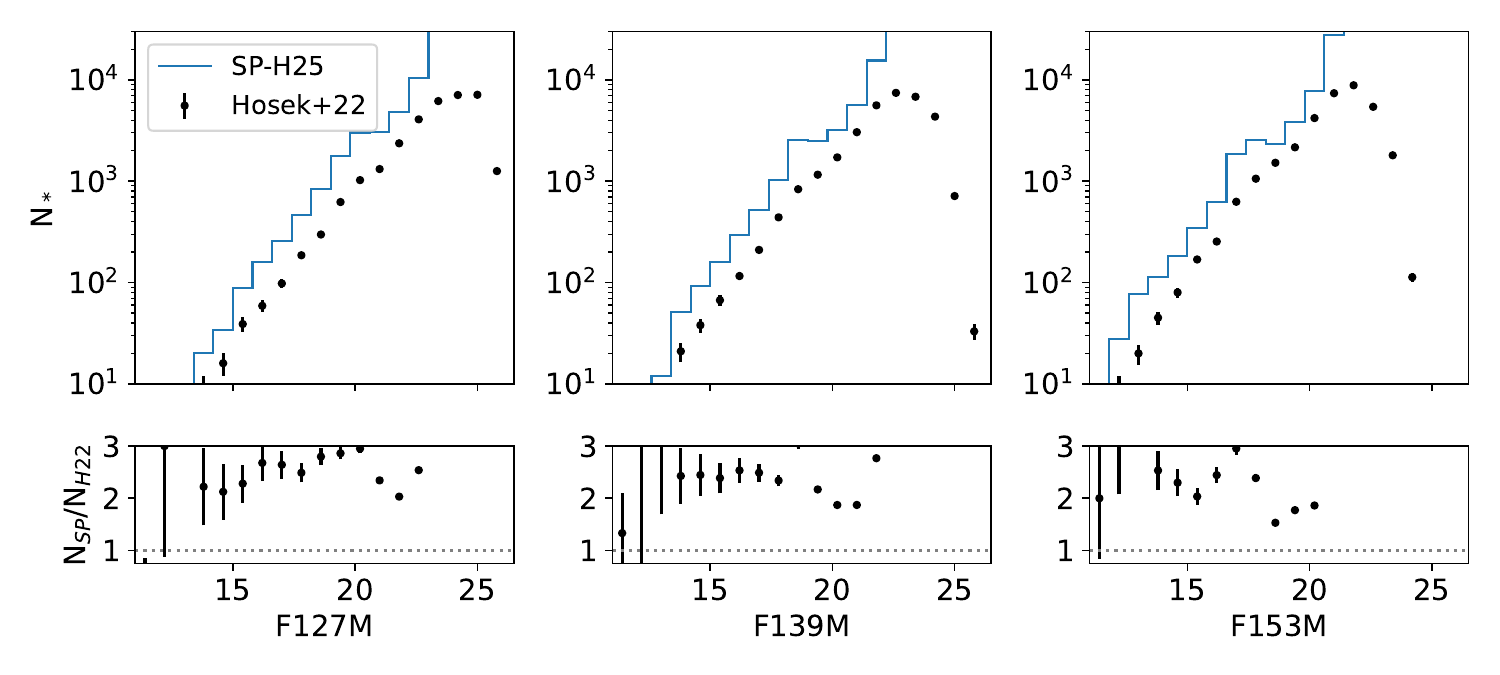}
    \caption{Luminosity functions from the SP-H25 model in comparison with the non-cluster members from the \citet{Hosek2022} Quintuplet (upper) and Arches (lower) field catalogs.}
    \label{fig:nsd_lf}
\end{figure}

\citet{Hosek2022} observed $2'\times2'$ fields in the direction of the Arches ($l=0.123^\circ$, $b=0.018^\circ$) and Quintuplet ($l=0.164^\circ$, $b=-0.060^\circ$) clusters with HST's WFC3-IR. From their catalogs, we filtered out cluster members (probability of cluster membership $>$ 0.5), as \synthpop does not currently model individual star clusters throughout the Galaxy. 
Figure \ref{fig:nsd_lf} shows the comparison of observed and simulated luminosity functions.
Again, we see very significant deviations between the SP-H25 populations and the observed LFs. In the Quintuplet field, the overprediction factor is typically $\sim0-50$\%, while the Arches field is often overpredicted by $\sim 100$\%. It is difficult to estimate how much incompleteness and Arches/Quintuplet contamination may contribute to these differences.
Still, the deviations call for a significant re-examination of the model at $|b|<0.5^\circ$, as discussed further in \S\ref{sec:disc}. 

Ultimately, without artificial star tests to account for incompleteness in many of these observational data sets, it is difficult to draw conclusions about model accuracy. In order to extract the full scientific potential of the upcoming Roman surveys of the Galactic bulge and plane (including their nuclear fields), such analyses will be necessary.

\section{Kinematic Model Evaluations}\label{sec:kine}

\subsection{The WFC3 Galactic Bulge Treasury Program}
\begin{figure}
    \centering
    \includegraphics[width=\linewidth]{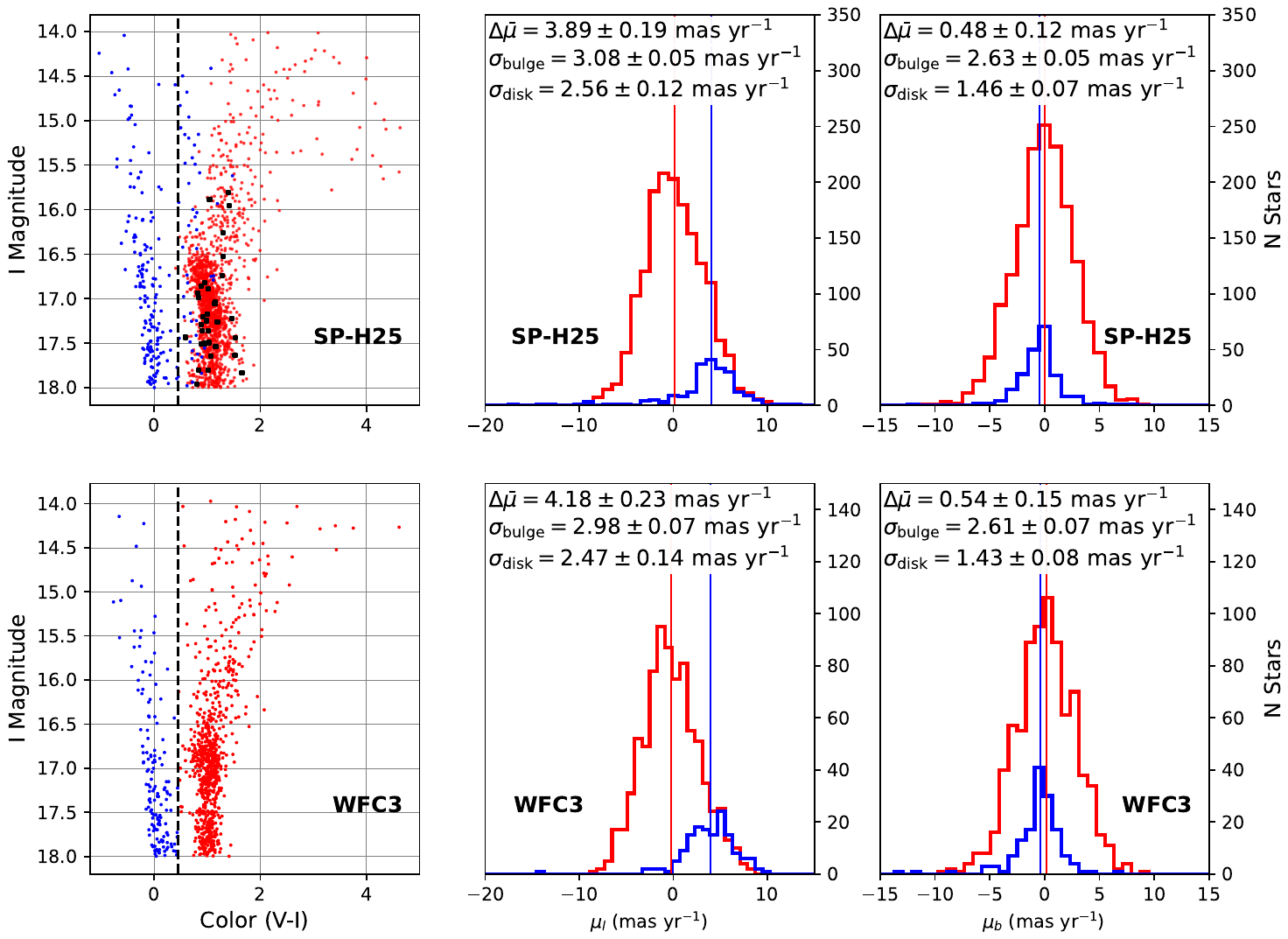}
    \caption{(Left two panels) CMDs used to distinguish red (bulge) and blue (disk) stars in the Stanek Window by color. In the top panel, we show SynthPop data and color code the model's  bulge (red points),  near disk (blue points) and far disk (black points) populations. In the WFC3 CMD, color coding simply denotes the position relative to a $(V-I)=0.45$ color cut to separate the populations, which is marked by a black dashed line. (Right four panels) Proper motion histograms for the selected red bulge stars and blue disk stars in Galactic longitude $\mu_l$ (middle column) and latitude $\mu_b$ (far column), respectively, for the SP-H25 model (top panels) and WFC3 data (bottom panels). Vertical lines show the mean of each distribution, and the mean differences and dispersions are shown in the top of each panel.}
    \label{fig:Stanekpm}
\end{figure}

As discussed in \S\ref{sec:bulge_treasury}, the Hubble WFC3 Galactic Bulge Treasury Program \citep{Brown2009,Brown2010} included the Stanek Window (l, b = [0.25, \textminus2.15]) which is within a few tenths of a degree of the RGBTDS survey region. This survey also measured proper motions. \citet{Terry2020} used this data to study the kinematics of the bulge, and we follow a similar procedure to separate the bulge and disk populations.
We apply a magnitude limit of 18 to $I$-band ST magnitudes for observed and simulated stars, as well as a bright star cut at 14 to the simulation catalog to better match the sensitivity of the observed population. 
We separate the red (bulge) and blue (disk) star populations using a color cut where red stars are $(V-I) > 0.45$ for both catalogs, as shown in Figure \ref{fig:Stanekpm}. In the SP-H25 catalog, we can distinguish near- and far- disk stars based on their distance and note that the ``disk stars'' in this analysis are near-disk stars, while far-disk stars get lumped into the ``bulge'' population but contribute minimally (\textless\ 3\%).

The WFC3 proper motions are provided as relative pixel position shifts in a reference frame where the mean proper motion is approximately zero. We apply the plate scale and time between observations to convert these to mas/yr relative proper motions.
We convert the SP-H25 absolute proper motions into a similar relative frame by subtracting the median value in the bulge – a reasonably stationary zero-point compared to the disk.
Figure \ref{fig:Stanekpm} shows the proper motion distributions of the color-separated populations. 

Because these are relative proper motions, we consider only the dispersions and the differences between the mean proper motions of the two populations, and not the absolute values of those means. We find the means and standard deviations of the distributions of disk and bulge proper motions and find differences in the means of [$\Delta\bar{\mu_l}$, $\Delta\bar{\mu_b}$ ] = [4.18 $\pm$ 0.23, 0.54 $\pm$ 0.15] mas yr$^{-1}$ for the WFC3 data and [$\Delta\bar{\mu_l}$, $\Delta\bar{\mu_b}$ ] = [3.89 $\pm$ 0.19, 0.48 $\pm$ 0.12] mas yr$^{-1}$ for the SynthPop catalog. For WFC3, we find $[\sigma_{bulge}, \sigma_{disk}]=[2.98\pm0.07, 2.47\pm 0.14]$ in $\mu_l$ and $[\sigma_{bulge}, \sigma_{disk}]=[2.61\pm0.07, 1.43\pm 0.08]$ in $\mu_b$ while, for SP-H25, we get $[\sigma_{bulge}, \sigma_{disk}]=[3.08\pm0.05, 2.56\pm 0.12]$ in $\mu_l$ and $[\sigma_{bulge}, \sigma_{disk}]=[2.63\pm0.05, 1.46\pm 0.07]$ in $\mu_b$. The simulated and observed values agree within $1.2\sigma$ in $l$ and $<1\sigma$ in $b$.
The velocity dispersions $\sigma_l$ agree within $1.2\sigma$ for the bulge and $0.5\sigma$ for the disk, while $\sigma_b$ is within $0.3\sigma$ for both the bulge and disk. Altogether, we consider this good agreement in the field nearest to the RGBTDS.

We perform the same analysis on the three remaining fields and provide the results in Table \ref{tab:wfc3_bulge_pm}. The agreement between simulated and observed values is reasonable for the second-nearest field to the Roman fields, SWEEPS, where all values agree within $<4\sigma$. The Baade's Window and the OGLE29 field also show a good agreement at $<2\sigma$. 

\begin{deluxetable}{lll}
\tablecaption{Proper motion offsets and dispersions computed for the Baade's Window, OGLE29, and SWEEPS fields using the same color-separation method as the Stanek Window shown in Figure \ref{fig:Stanekpm}.    \label{tab:wfc3_bulge_pm}}
\tablehead{\colhead{Field} & \colhead{SP-H25} & \colhead{WFC3}}
    \startdata
        \begin{tabular}{l}
        Baade's Window \\
        $(l, b) =[1.06, -3.81]$ 
        \end{tabular}
        &
        \begin{tabular}{l}
        $[\Delta\mu_l, \Delta\mu_b] = [2.63\pm0.33, 0.61\pm0.22]$ \\
        $[\sigma_{\mathrm{bulge}}, \sigma_{\mathrm{disk}}]_l = [2.73\pm0.07, 3.26\pm0.22]$ \\
        $[\sigma_{\mathrm{bulge}}, \sigma_{\mathrm{disk}}]_b = [2.63\pm0.07, 1.99\pm0.14]$
        \end{tabular}
        &
        \begin{tabular}{l}
        $[\Delta\mu_l, \Delta\mu_b] = [3.36\pm0.33, 0.52\pm0.25]$ \\
        $[\sigma_{\mathrm{bulge}}, \sigma_{\mathrm{disk}}]_l = [2.86\pm0.10, 2.96\pm0.21]$ \\
        $[\sigma_{\mathrm{bulge}}, \sigma_{\mathrm{disk}}]_b = [2.56\pm0.09, 2.10\pm0.15]$
        \end{tabular}
        \\
        \tableline
        \begin{tabular}{l}
        OGLE29 Field \\
        $(l, b) =[-6.75, -4.72]$ 
        \end{tabular}
        &
        \begin{tabular}{l}
        $[\Delta\mu_l, \Delta\mu_b] = [3.20\pm0.38, 0.82\pm0.26]$ \\
        $[\sigma_{\mathrm{bulge}}, \sigma_{\mathrm{disk}}]_l = [2.35\pm0.10, 3.22\pm0.25]$ \\
        $[\sigma_{\mathrm{bulge}}, \sigma_{\mathrm{disk}}]_b = [1.84\pm0.08, 2.20\pm0.17]$
        \end{tabular}
        &
        \begin{tabular}{l}
        $[\Delta\mu_l, \Delta\mu_b] = [2.53\pm0.46, 0.41\pm0.44]$ \\
        $[\sigma_{\mathrm{bulge}}, \sigma_{\mathrm{disk}}]_l = [2.32\pm0.12, 3.11\pm0.31]$ \\
        $[\sigma_{\mathrm{bulge}}, \sigma_{\mathrm{disk}}]_b = [2.00\pm0.11, 2.90\pm0.29]$
        \end{tabular}
        \\
        \tableline
        \begin{tabular}{l}
        SWEEPS Field  \\
        $(l, b) =[1.26, -2.65]$ 
        \end{tabular}
        &
        \begin{tabular}{l}
        $[\Delta\mu_l, \Delta\mu_b] = [3.43\pm0.22, 0.48\pm0.15]$ \\
        $[\sigma_{\mathrm{bulge}}, \sigma_{\mathrm{disk}}]_l = [3.12\pm0.06, 2.65\pm0.15]$ \\
        $[\sigma_{\mathrm{bulge}}, \sigma_{\mathrm{disk}}]_b = [2.73\pm0.05, 1.61\pm0.09]$
        \end{tabular}
        &
        \begin{tabular}{l}
        $[\Delta\mu_l, \Delta\mu_b] = [3.06\pm0.38, 0.64\pm0.25]$ \\
        $[\sigma_{\mathrm{bulge}}, \sigma_{\mathrm{disk}}]_l = [3.08\pm0.11, 3.48\pm0.25]$ \\
        $[\sigma_{\mathrm{bulge}}, \sigma_{\mathrm{disk}}]_b = [2.51\pm0.09, 2.19\pm0.16]$
        \end{tabular}
        \\
    \enddata
\end{deluxetable}

\subsection{Gaia \& VVV Proper Motions}\label{sec:gaia_pm}

\begin{figure}
    \centering
    \includegraphics[width=\linewidth, trim={0 0.4in 0 0.8in},clip]{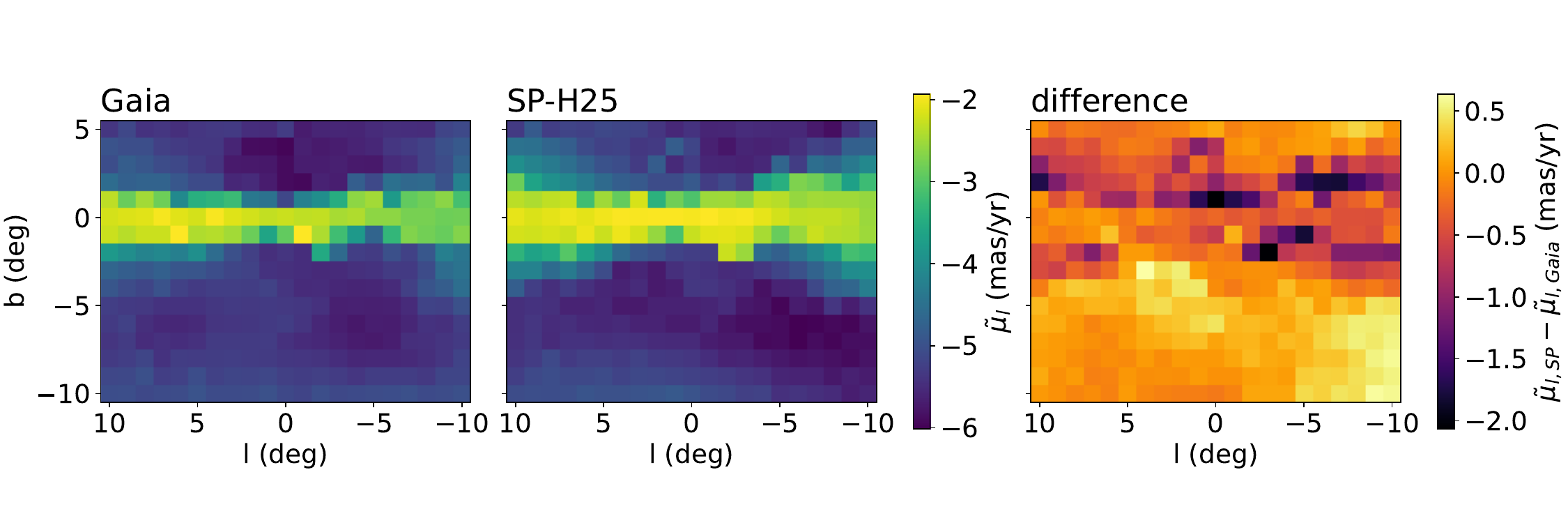}
    \includegraphics[width=\linewidth, trim={0 0.4in 0 0.8in},clip]{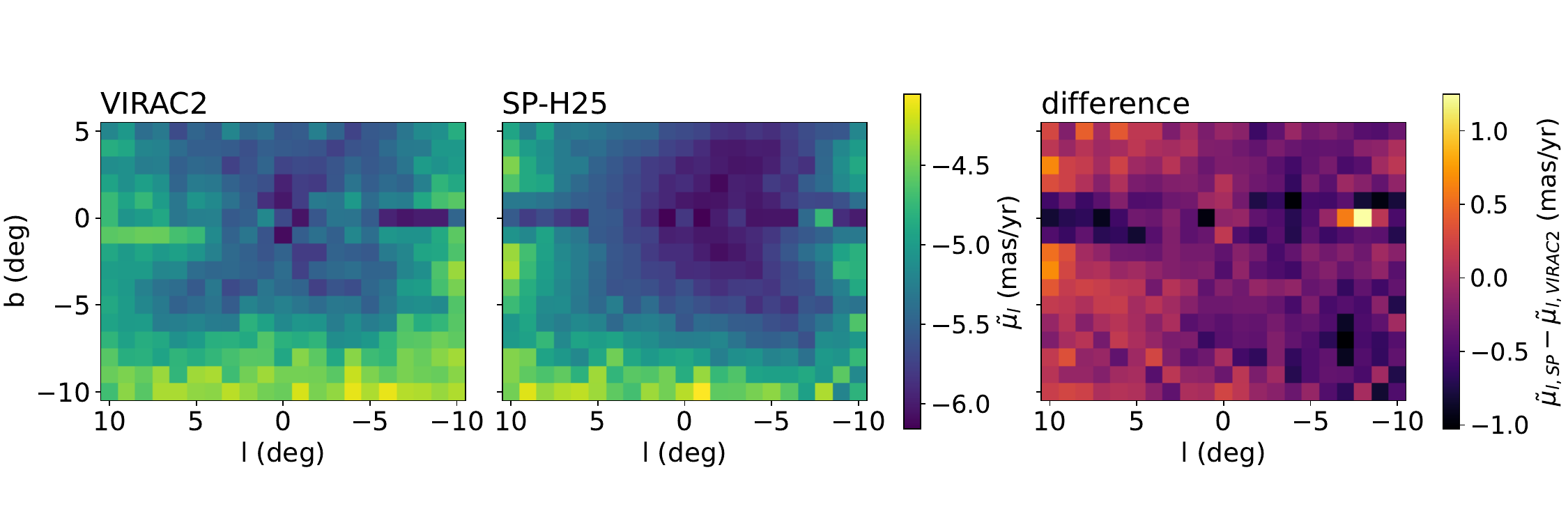}
    \includegraphics[width=\linewidth]{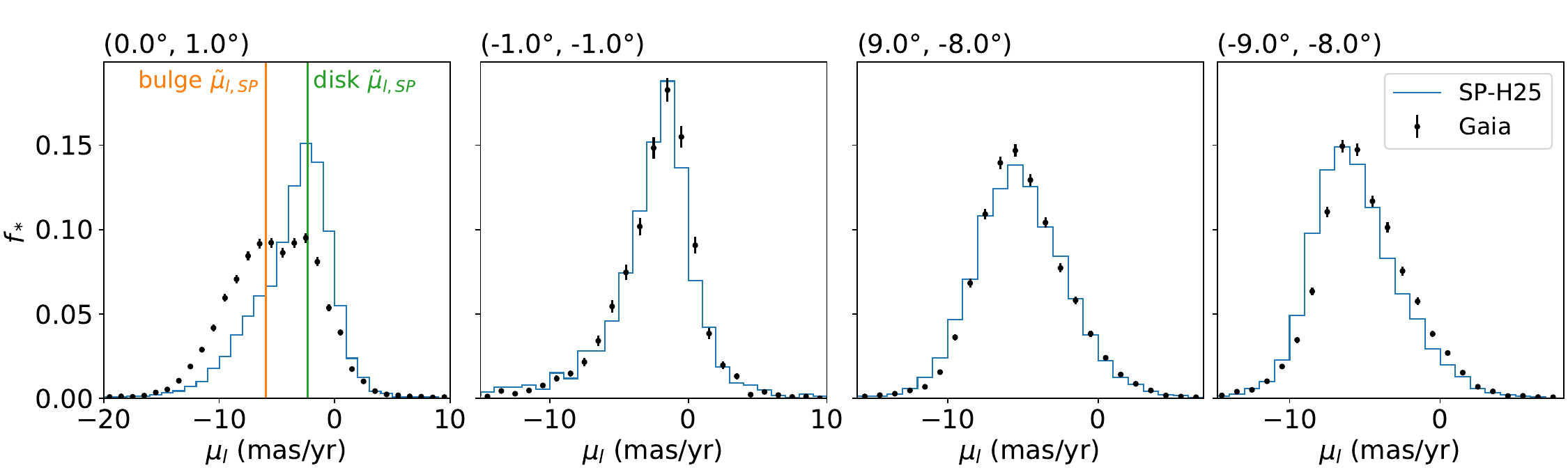}
    \caption{Proper motion component $\mu_l$ comparison between observational catalogs and SP-H25. The top row shows the median values for Gaia DR3 in a grid across the sky, as well as their difference. The middle row shows the same for VIRAC2. The bottom row shows Gaia $\mu_l$ histograms for four individual sightlines that cover different ends of the spectrum for the systematic trends in the upper panel.}
    \label{fig:gaia_mul}
\end{figure}

Gaia Data Release 3 \citep[DR3][]{GaiaCollaboration2016,GaiaCollaboration2023} provides an extensive catalog of stars with well-measured proper motions across the sky. Figure \ref{fig:gaia_mul} shows how the distributions of $\mu_l$ compare to those simulated via SynthPop with $1^\circ$ grid sampling across the region of sky covered by our extinction map. 
It is important to note when examining Gaia DR3 proper motions that Gaia's astrometric performance worsens in crowded fields, and proper motion uncertainties in the bulge fields are underestimated \citep{Luna2023}. 
In order to better peer through the strong extinction in the plane, we also examine proper motion measurements from the VIRAC2 catalog, which reported $\sim$0.37 mas yr$^{-1}$ precisions for $11\leq Ks\leq 14$ and $\sim$1.5 mas yr$^{-1}$ at Ks$=$16 \citep{Smith2025}. For the SP-H25 catalogs we compare to VIRAC2, we again use a blend radius of 0.36\arcsec and compute Ks-band flux-weighted mean proper motions for each blend. For Gaia, we limit the sample to stars with G$<$ 20 mag, and we limit VIRAC2 to K$_{\rm s}<16$.

The residual difference between the median SP-H25 and Gaia $\mu_l$ shows clumpy structure within $|b|\lesssim3^\circ$. The extinction map used in this work is based on bulge stars and uses a disk-scaling scheme. Thus, it may perform poorly for optical simulations of nearby disk stars in regions of high extinction, particularly where that extinction is irregularly distributed in the near disk. 
To confirm this explanation for the disagreement at these low absolute latitudes, we examined extinction along several sample sightlines in the model used by SP-H25 against a 3-d extinction map calculated from nearby disk stars. We use the \citet{Lallement2022} map for this comparison (using our extinction law to convert their extinction at 5500 \AA~ to $\rm A_{Ks}$), which used Gaia EDR3 parallaxes and Gaia + 2MASS photometry to estimate extinction as a function of position across a $6\times6\times0.8$ kpc$^3$ volume centered on the Sun. At an example sightline with good model-data $\mu_l$ agreement ($l=-1^\circ$, $b=-1^\circ$), the extinction maps agreed well along the line of sight, with a total extinction at 3 kpc of $A_{\rm Ks}=0.27$ in the \citet{Lallement2022} map and 0.28 in ours. At an example sightline with significant $\mu_l$ disagreement ($l=0^\circ$, $b=1^\circ$), the \citet{Lallement2022} map provides significantly higher extinction for nearby stars; at 3 kpc, $A_{\rm Ks}=0.36$ in the \citet{Lallement2022} map and only 0.15 in ours.
These larger differences in the extinction values between 0-3 kpc along the sightlines with this kinematic disagreement supports our hypothesis that our 3-d extinction model inaccuracies within a few kpc bias the sample of simulated Gaia stars used in the proper motion comparison.


In addition to the disagreement caused by extinction near the plane, we see a gradient in the level of $\mu_l$ model-data disagreement at low latitudes ($b\leq-5^\circ$), going from about 0.1 mas/yr at $l=10^\circ$ to about 0.5 mas/yr at $l=-10^\circ$. This may be due to an imperfect bulge kinematic model, or an inaccurate Galactic bar angle used in the bulge density model.
A similar trend exists in $\mu_b$ proper motions, where extinction produces clumpy variation in the plane, and the lower bulge shows a gradient; however, both are less dramatic in this direction as it is perpendicular to Galactic rotation.

In the VIRAC2 comparison, this dust issue is much less impactful. However, the SP-H25 catalog shows significant asymmetry across $l$, where the observational data shows instead a more clear and symmetric X-shaped structure. We again see a gradient in the model-data residual across the bulge as a function of longitude. Interestingly, this gradient with $|l|$ has the opposite sign as that in the Gaia data. The root cause of this disagreement is is not clear from this examination alone, and further study will be required to improve our model of the Galactic bulge's kinematics. The \iroman Galactic Plane Survey (GPS), which will cover $|b|<6^\circ$ for $|l|<10^\circ$, in addition to a thinner span in $b$ at larger $|l|$, will provide a deeper look at proper motions across much of this region. 


\subsection{Nuclear Stellar Disk Proper Motions}
\begin{figure}
    \centering
    \includegraphics[width=0.6\linewidth]{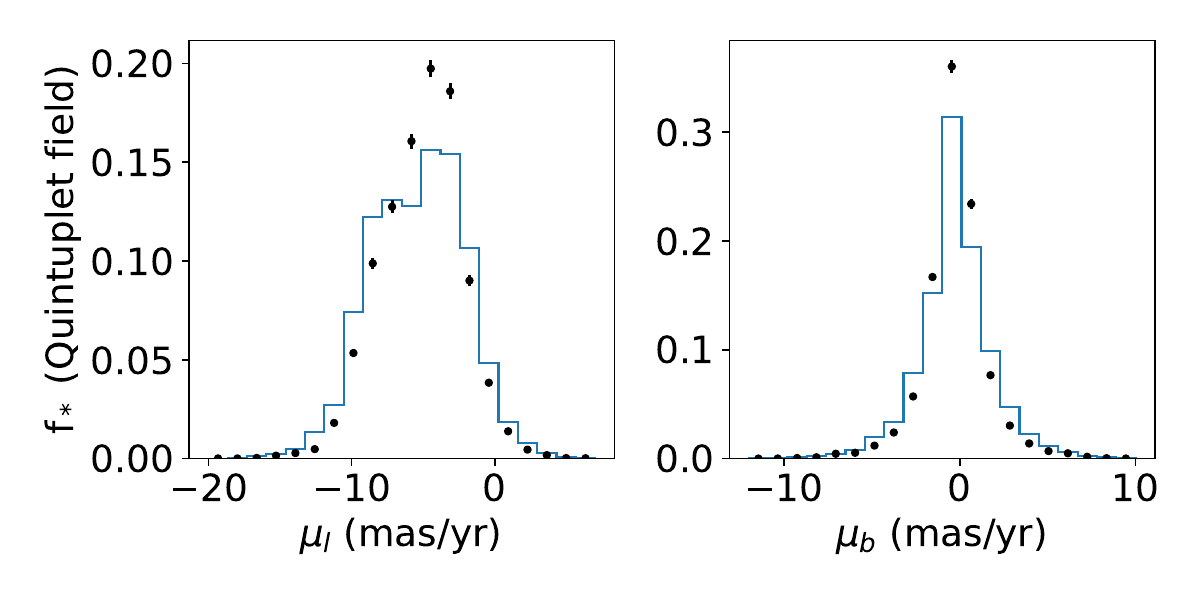}
    \includegraphics[width=0.6\linewidth]{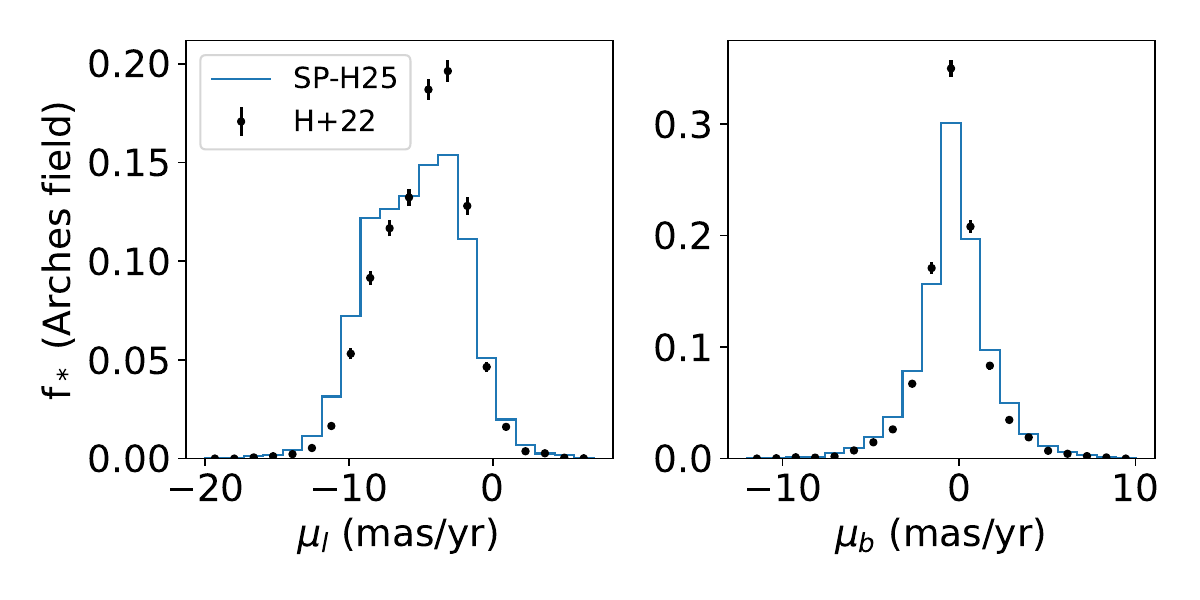}
    \caption{Proper motion comparison between the SP-H25 model and the non-cluster stars from the \citet{Hosek2022} Quintuplet (left) and Arches (right) cluster field catalogs. 
    }
    \label{fig:nsd_pms}
\end{figure}

The \citet{Hosek2022} catalog described in \S\ref{sec:nsd_lfs} also provides proper motions from F153M imaging. The proper motions are shown in Figure \ref{fig:nsd_pms} in comparison with SP-H25 simulated catalogs. We limit this sample to simulated and observed stars with F153M$<$20, beyond which the luminosity function comparison above suggests that completeness of the catalog begins to strongly decline. The observed proper motion distributions show tighter peaks, suggesting that the SP-H25 model tends to overpredict velocity dispersions near the Galactic Center. An improper balance of NSD and bulge stars may also contribute to this difference. The observed distribution in $\mu_l$ appears to have a stronger peak at $\sim$-3 mas/yr and a weaker shelf toward $\sim-7$ mas/yr than the SP-H25 model produces.

\subsection{BRAVA Radial Velocities}
The Bulge Radial Velocity Assay (BRAVA) spectroscopic survey measured the radial velocities (RVs) of $\sim10,000$ M-giant stars in the bulge \citep{Howard2008brava, Kunder2012brava}. 
Their targets were selected from the Two Micron All Sky Survey \citep[2MASS; ][]{Skrutskie20062mass}, with a magnitude selection of $9.25 > K_{\rm S} > 8.2$.
The fields primarily span $-10^{\circ} \leq l \leq +10^{\circ}$ and $-8^{\circ} \leq b \leq -4^{\circ}$, each with a 40\arcsec field of view.
For each observed field, they report the mean heliocentric and galactocentric RVs $\langle V_{\rm HC}\rangle$ and $\langle V_{\rm GC}\rangle$, the RV dispersion $\sigma$, and the distribution of galactocentric RVs in the field.
While radial velocities themselves do not impact microlensing surveys' predictions or interpretation, they can help probe the overall accuracy of the kinematic model. 

For each BRAVA field, we create SP-H25 catalogs with simulated 2MASS photometry. 
We select stars falling in the same $K_{\rm S}$ range used by the survey, and filter for stars generated from \synthpop's bulge population to exclude RVs from foreground stars in our calculations.
We fit a normal distribution to each catalog's RVs to obtain $\langle V_{\rm BC}\rangle$ and $\sigma$, with uncertainties $\frac{\sigma}{\sqrt{N}}$ and $\frac{\sigma}{\sqrt{2N}}$ respectively.

We find the \synthpop RVs to have a slightly steeper slope as a function of $l$ than the BRAVA RVs. 
The \synthpop mean velocities typically agree with the BRAVA results within 1-2$\sigma$, though three fields differ by as much as $\sim4.5\sigma$.
The model-data differences are strongest at negative $l$ values based on the best-fit slope. At the high $l$ end, the slope of $\langle V_{\rm BC} \rangle$ begins to turn over in both the model and data. At both extremes in $l$, agreement is worse for fields further from the Galactic plane ($-8^\circ\lesssim b \lesssim-5^\circ$).
The \synthpop velocity dispersions are typically within 1-2$\sigma$ of the BRAVA results. The RGBTDS fields lie in the inner region where agreement is better. 

\begin{figure}
    \centering
    \includegraphics[width=\linewidth]{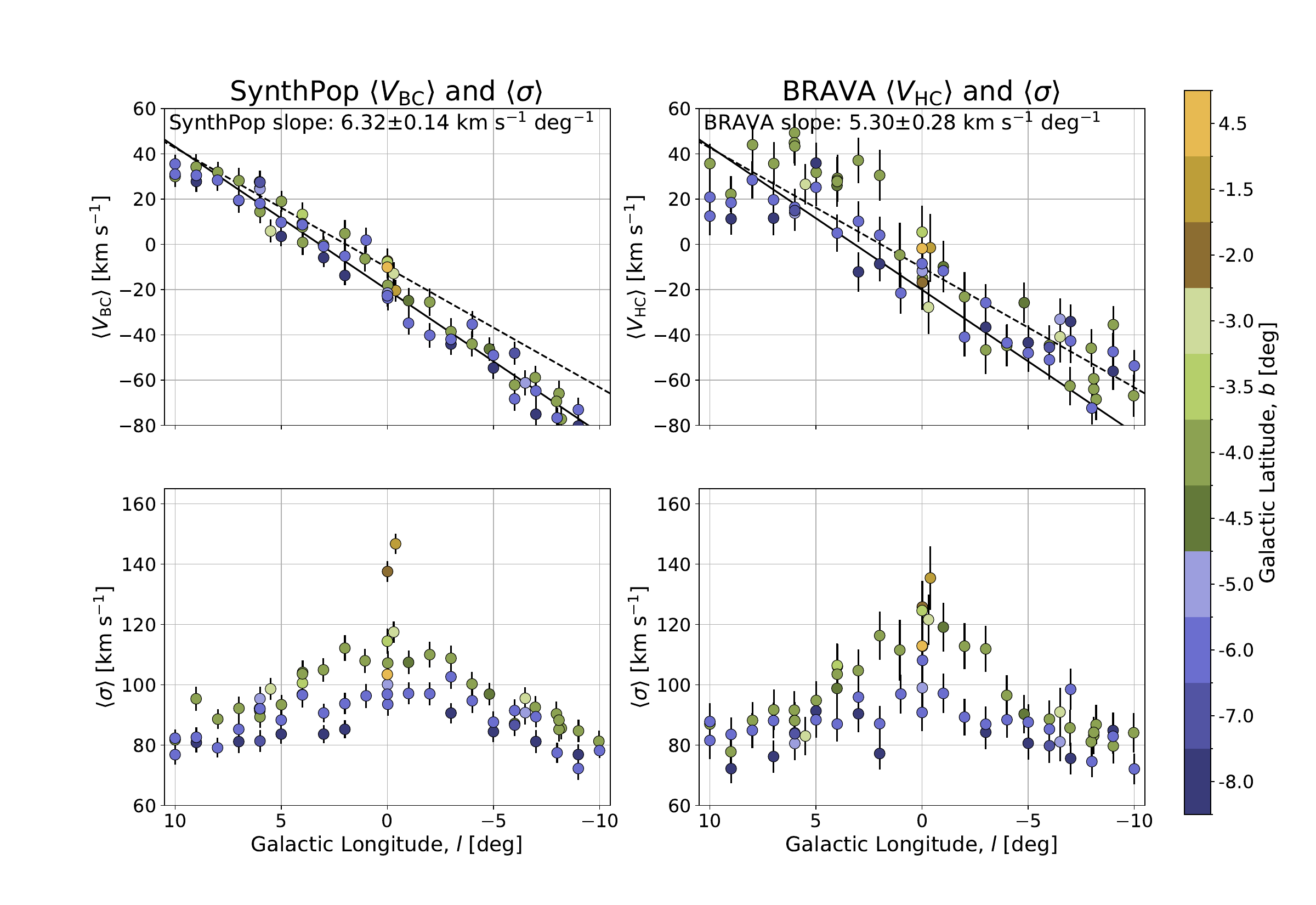}
    \caption{The mean radial velocities (top row) and mean velocity dispersions (bottom row) of simulated SP-H25 catalogs compared to the observed values from the BRAVA survey. Each color represents a different galactic latitude, $b$. In the mean RV panels, the solid black lines represent the best fit line of the \synthpop data, while the dashed black lines represent the best fit line of the BRAVA data. The slope of each is given at the top of the respective panels.}
    \label{fig:rvs}
\end{figure}

\section{Microlensing Model Evaluations}\label{sec:mulens}

Microlensing event rates, optical depths, and timescale distributions provide a valuable test of Galactic models, particularly when seeking to use the model for further microlensing simulations, because they depend on all aspects of the Galaxy's structure and kinematics in order to predict stellar positions, magnitudes, masses, and relative motions.

Two key properties of each potential microlensing event are the physical scale of the event (the Einstein ring, $\theta_E$), and the timescale of the event ($t_E=\theta_E / \mu_{\rm rel}$, where $\mu_{\rm rel}$ is the relative proper motion between the lens and source stars). 
The angular Einstein ring radius is calculated as: 
\begin{equation}
    \theta_E = \left(\frac{4 G M_l}{D_{rel}c^2}\right)^{1/2},
\end{equation}
where $G$ is the gravitational constant, $M_l$ is the mass of the lens star, and $c$ is the speed of light. $D_{rel}$ is defined as $(D_l^{-1}-D_s^{-1})^{-1}$, where $D_l$ and $D_s$ are the distances from the Sun to the lens and source, respectively. 

\subsection{Monte Carlo Integration}
\label{sec:eqs}

There are four microlensing rates/distributions commonly characterized by observations in the literature: event rate per unit area, event rate per source star, optical depth, and average timescale. These are typically computed as average observed values within spatial bins. With a Galactic model, we can generate pencil-beam catalogs at the set bin spacing and perform Monte Carlo integration to draw many possible lens-source pairs and simulate the observable quantities. For these sums, we consider all possible lens-source pairs from our catalogs, selecting events where the baseline magnitude of the source meets some threshold, the source is more distant from us than the lens, and the event timescale is within some range selected to match a given survey. These are intrinsic underlying event rates, and completeness corrections and detection efficiencies are required for observational results to be properly compared.

Microlensing optical depth, or the fraction of sky covered by lens stars' Einstein rings, ($\overline{\tau}$), is:
\begin{equation}
\overline{\tau} = \frac{1}{N_s} \sum_{}^{N_s} \left( \frac{\pi}{\Omega_l}  \sum_{D_s> D_l}^{N_l} \theta_E^2 \right),
\end{equation}
where $N_s$ is the number of source stars, $\Omega_l$ is the solid angle of sky used for the lens catalog simulations, and $N_l$ is the number of lens stars. The summation iterates over each possible source star (which must be brighter than some limiting magnitude) and each of its possible lens stars (any star less distant than the source). We note here that Einstein rings are dependent on the presence of observable sources past each given lens star, so average optical depth is not simply a function of stellar masses and number densities, but also on stellar magnitudes and thus extinction.

The average rate of microlensing events per source star, $\overline{\Gamma}_{*}$, is:
\begin{equation}
	\overline{\Gamma}_{*} = \frac{1}{\Omega_l N_s} \sum^{N_s} \sum_{D_s>D_l}^{N_l} 2 \mu_{\rm rel} \theta_E.
\end{equation}
We also consider event rate in terms of angular area on-sky, $\Gamma_{\Omega}$:
\begin{equation}
	\Gamma_{\Omega} = \frac{1}{\Omega_l \Omega_s} \sum^{N_s} \sum_{D_s>D_l}^{N_l} 2 \mu_{\rm rel} \theta_E
    = \frac{N_s}{\Omega_s} \overline{\Gamma}_{*},
\end{equation}
where $\Omega_s$ is the solid angle of the source star catalog simulation. The average timescale for a patch of sky, $\overline{t_E}$, is calculated as:
\begin{equation}
\overline{t_E} = \frac{\sum^{N_s} \sum_{D_s>D_l}^{N_l} t_E * \theta_E \mu_{rel}}{\sum^{N_s} \sum_{D_s>D_l}^{N_l} \theta_E \mu_{rel}}.
\end{equation}
This is a weighted mean where possible events are weighted by $\theta_{\rm E} \mu_{\rm rel}$, or the rate of solid angle swept out by the lens-source pair's Einstein ring. See Appendix \ref{app:weights} for further discussion of these equations in comparison to other microlensing rate modeling methods. 

\subsection{Optical Surveys}\label{sec:ogle}

Galactic models have long faced challenges fitting microlensing data \citep[see e.g.][]{Kerins2009, Sumi2013}, though part of the issue has been shortcomings in the analysis of microlensing survey data itself, usually in the form of assumptions that did not hold. Early observational studies focused on measurements of the optical depth because it is, in theory, independent of kinematics and can be computed with a simple integral just through the density distribution of the Galaxy. However, practical microlensing survey analyses are limited to measurable event timescales (which do depend on kinematics), and the optical depth contribution of each event is proportional to its timescale. Additionally, each event's contribution to the Poisson uncertainty adds in quadrature proportional to $t_{\rm E}^2$ \citep{1995ApJ...449..521H}, so the measurement is quite sensitive to the timescale range studied. This led to a greater emphasis being placed on the microlensing event rate per star, \citep[e.g.,][]{Sumi2013}, which has smaller Poisson errors but might be more sensitive to evaluation of the detection efficiency for short events. 

When planning for Roman was in its early stages, the discrepancy between models and data became a significant issue, with models apparently significantly underestimating event rates~\citep[e.g.,][]{Green2012,Spergel2013,Penny2013}. \citet{Sumi2016} recognized that a key assumption of \citet{Sumi2013}, that red clump star counts (which were used for normalization of the luminosity function), would be complete, was incorrect and had led to systematic overestimates of the event rate (and optical depth). \citet{Sumi2016} made a quick correction to the \citet{Sumi2013} results by comparing MOA red clump counts to OGLE counts from \citet{Nataf2013}, but full artificial star tests are needed to fully account for completeness, especially in ``all-star'' event rate studies that use all events found in a difference imaging search. 

More recent optical microlensing event rate measurements by \citet{Mroz2019} using OGLE data applied completeness corrections to their luminosity functions and found lower event rates than even \citet{Sumi2016}. \citet{Specht2020} performed a Galactic model-based analysis of the OGLE-IV results using the Besan{\c c}on model version 1307 \citep[see also][]{Robin2014,Awiphan2016} in the second-generation Manchester-Besan{\c c}on Microlensing Simulator (MaB$\mu$lS-2), finding that the model improved on past work but still under-predicts event rate per unit area. We note that \citet{Specht2020} results included an error in the MaB$\mu$lS weighting scheme for event timescales, but they agreed well with observation due to tuning of the low-mass/brown dwarf IMF extension (see Appendix \ref{app:weights}).
The most recent analysis of MOA data by \cite{Nunota2025} found that photometric calibration also played a role in inflating event rates, and now with both issues resolved, MOA and OGLE event rates are consistent with each other. \citeauthor{Nunota2025} also analyzed their results in comparison to the MaB$\mu$lS-2 \citep{Specht2020} and {\tt genulens} \citep{Koshimoto2022} simulations, finding better agreement with {\tt genulens} but discrepancies with both in the $-3^\circ \leq b \leq -1^\circ$ region and no observational coverage for $b>-1^\circ$.

\begin{figure}
    \centering
    \includegraphics[width=0.9\linewidth]{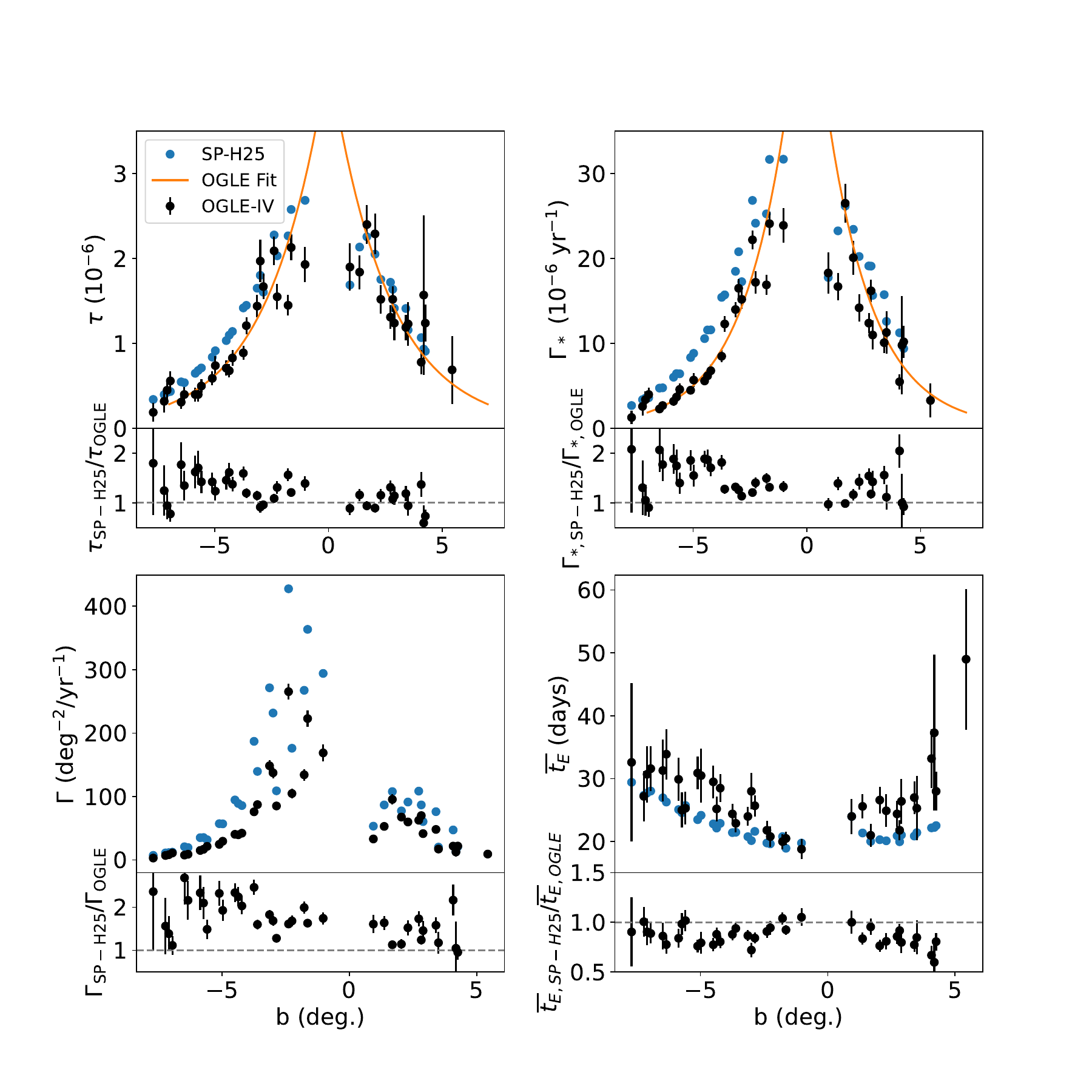}
    \caption{OGLE-IV survey event rates and distributions for $t_E<300$ day events as a function of Galactic latitude in comparison to SP-H25 model simulations, with a residual panel below each showing the model/observed value ratios. The top row shows microlensing optical depth (left) and event rate per star (right), with the best-fit lines to OGLE-IV data \citep{Mroz2019} included. The bottom row shows the event rate per square degree (left) and average timescale (right).}
    \label{fig:ogle-bdep}
\end{figure}

Here, we compare our Galactic model to these latest completeness-corrected OGLE event rates toward the Galactic bulge \citep{Mroz2019}, which cover a wider area of sky than the agreeing MOA results. We apply the equations in \S\ref{sec:eqs} to synthetic catalogs from the SP-H25 model to generate microlensing observables for the OGLE-IV fields with a source star magnitude limit of $I\leq21$ and timescales limited to $t_E\leq 300$ days. Figure \ref{fig:ogle-bdep} shows the results as a function of Galactic latitude for OGLE-IV fields within $|l|<3^\circ$ with $>$1 microlensing event detected. 

The SP-H25 model overpredicts microlensing optical depth by a factor of $\sim1.1$, event rate per source star by $\sim1.2$, event rate per unit area by $\sim1.5$, and underpredicts average timescale by $\sim0.9$. The effect tends to be more severe south of the Galactic plane, except for timescale. It is unsurprising that a model that excludes stellar multiplicity would consistently underpredict typical microlensing timescales, as larger system masses (and non-zero separations) lead to longer $t_E$ \citep[see e.g.][]{Abrams2025}. This effect may also contribute to the over-prediction of the other observables, as a higher mass per system with an unchanged Galactic stellar mass distribution produces fewer star systems and thus likely fewer random chance stellar crossing events, though the extent of the effect depends on the companion mass ratio distribution and orbital parameters. Additionally, issues like proper motion distributions, stellar density, and extinction contribute as well, but these cannot be disentangled from microlensing observables alone.

In addition model shortcomings, issues in microlensing event statistics from observation may also still contribute to the disagreement, despite the resolution of several past issues discussed above. \citet{Mroz2019} fit their events to point-source point-lens models that included blending but not microlensing parallax, multiplicity, or finite source effects. They noted that 10\% of events detected by their search algorithm were ``anomalous'' (events with clear binary lens features or strong microlensing parallax) and excluded from the event rate calculation, to which they multiplied a correction factor of 1.09. They noted that they likely overestimate detection efficiency of long-timescale events as annual microlensing parallax was excluded from their injected events. While their event fits included blending, but it can be difficult to disentangle correlations between impact parameter, blending, and timescale for highly blended events \citep{Wozniak1997}. Finite source effects may also bias measurements of impact parameter and timescale. \citet{Abrams2025} found that 52\% of microlensing events in their simulation that were fit well with point-source point-lens model (also with no parallax included) actually contained a binary lens or source system (or both). This suggests that despite filtering out obvious binary events, \citet{Mroz2019} likely include many binary events in their sample. Thus, a Galactic model without stellar multiplicity should not be expected to perfectly reproduce their result.

\subsection{Infrared Surveys} \label{sec:ir_surveys}
Because the primary long-term microlensing programs have been optical, IR microlensing rates, like those the RGBTDS will see, are not as well characterized. We explore the IR microlensing event rates available, but note that this analysis is severely limited by a lack of detection efficiency characterization for the available surveys. 
While the observed optical event rates presented in \S\ref{sec:ogle} have already been corrected for detection efficiency and the effect of blending on the number of source stars observed, the observations in this section have not. Thus, we estimate the effect of blending in our simulated per star event rates by converting from underlying star counts to apparent star counts, as they are estimated following the blending scheme described in \S\ref{sec:counts}. Detection efficiency handling varies between the two data sets depending on what was available, as described below.

\citet{Wen2023} presented raw event rates from the UKIRT microlensing survey with no detection efficiency corrections, and thus they are considered lower limits on the true event rate per star. The survey operated in both H and K filters, with the primary filter varying across fields. The H-band magnitude range\footnote{\url{https://exoplanetarchive.ipac.caltech.edu/docs/UKIRT_figures.html}} is 11.5-19.0, and K-band is 11.5-18.0. We use the method described in \S\ref{sec:eqs} to estimate event rates in each of these filters for source stars within their magnitude limits with simulation fields sampled at 0.1$^\circ$ spacing between $-1^\circ<l<1^\circ$ and $-2.0^\circ<b<0.5^\circ$. We correct the number of sources using the blending scheme described above with a blend radius of 0.225\arcsec. Additionally, the UKIRT event catalog contains events with $|u_0|\lesssim1.5$, so we scale our event rates up from the $|u_0|\leq1$ values accordingly. We limit our simulation $t_E$ range to that of UKIRT's detected events: 1.5-350 days.

\begin{figure}
    \centering
    \includegraphics[width=0.5\linewidth]{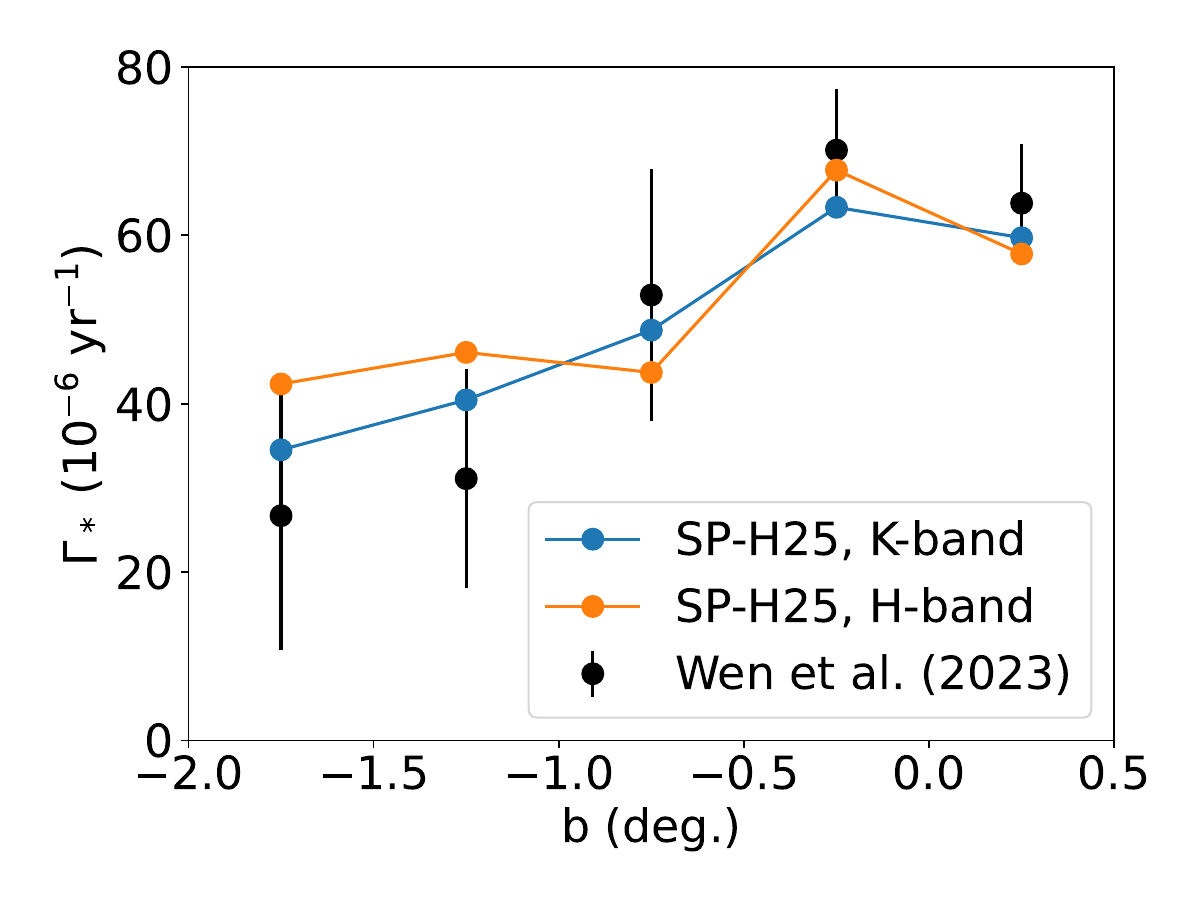}
    \caption{Microlensing event rate per star from the \citet{Wen2023} UKIRT raw event rates, which are not corrected for detection efficiency and should be interpreted as lower limits. We show K- and H- band simulated rates for $1.5\leq t_E\leq 350$ day events with the SP-H25 model, as some survey patches used each as their primary filters.}
    \label{fig:ukirt}
\end{figure}

Our results are shown in comparison with the \citet{Wen2023} results in Figure \ref{fig:ukirt}. 
The SP-H25 model agrees with these lower limit values within the (relatively large) observational uncertainties. They model may, in fact, be an under-prediction of event rates, since a detection efficiency-corrected UKIRT event rate would be higher by an unknown factor. Ultimately, a corrected event rate examination would be required to draw meaningful conclusions.

While the VVV survey was not designed for microlensing, it detected 1959 microlensing events \citep{Husseiniova2021}, in the direction of the Galactic bulge and plane between 2010-2019. \citet{Navarro2020} examined a separate sample of 238 events detected in VVV between 2010-2015 within $-1.25^\circ < l < 0.26^\circ$ and $-3.7 < b < 3.9$. \citeauthor{Navarro2020} noted that in Ks-band, in contrast to the optical microlensing surveys, the efficiency-corrected event rate continues to increase with decreasing $|b|$ all the way to $0^\circ$.

\citet{Husseiniova2021} computed the microlensing detection efficiency of their pipeline for 5 discrete $t_E$ values from 3 to 300 days for $0.5^\circ\times0.5^\circ$ spatial bins by injecting microlensing events into real VVV light curves. The simulated microlensing events had baseline magnitudes distributed according to the observed star catalogs, uniformly distributed impact parameters ($u_0$) and uniformly distributed blending parameters (the ratio of source flux to combined lens-source-neighbor flux, or source flux fraction). We use their event catalog and detection efficiency tables to estimate event rates. 

We examine the latitude dependence of these near-IR event rates as these are relevant to the RGBTDS and can be directly compared to the UKIRT rates. Thus, we separate the events into latitude bins with $1^\circ$ spacing in $b$ with longitudes limited to $|l|<4^\circ$ for the full extent of $b$ covered by our extinction map. Within a given spatial bin $i$, we average the \citet{Husseiniova2021} detection efficiencies into one $\epsilon_{tE,i}$ for each timescale bin $t_{\rm E}$. Using all detected events $j$ in the spatial bin $i$, we calculate the event rate:
\begin{equation}
\Gamma_* = \frac{1}{N_{*,i} \Delta T}\sum_j\frac{1}{\epsilon_{t_{\rm E},i}},
\end{equation}
where $N_{*,i}$ is the number of catalog stars in the spatial bin and $\Delta T$ is the survey duration. We estimate uncertainties for all bins based on Poisson uncertainties for the event count in each $t_{\rm E}$ bin and, for $t_E$ bins with no events, adopt a $+1\sigma$ Poisson uncertainty of 1.

\begin{figure}
    \centering
    \includegraphics[width=0.5\linewidth]{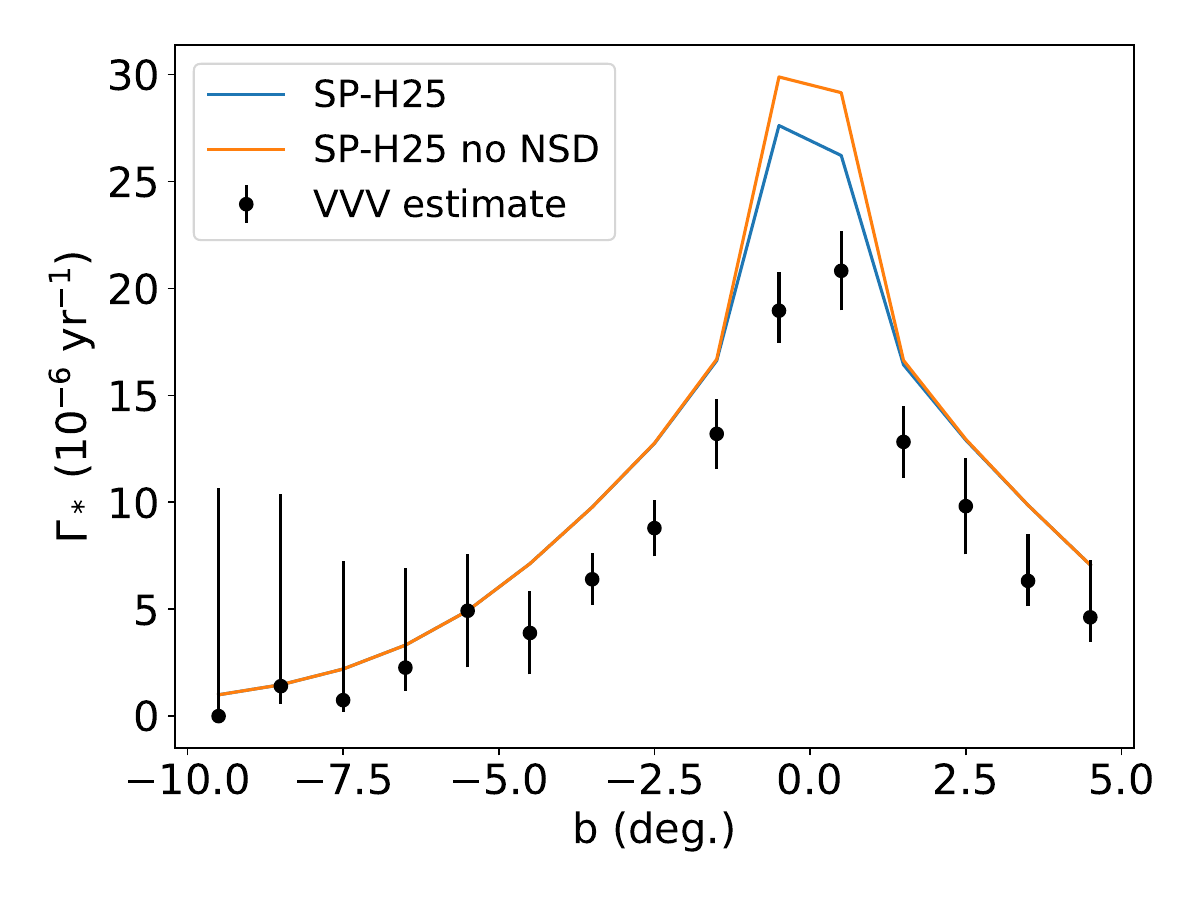}
    \includegraphics[width=0.4\linewidth]{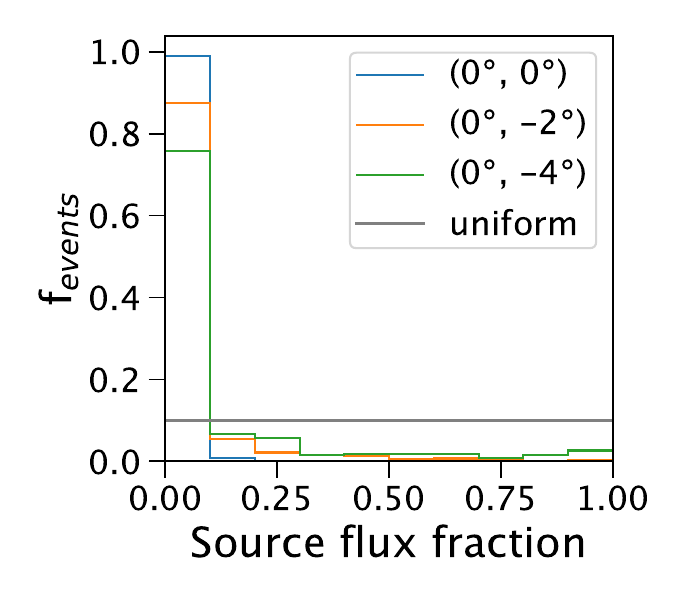}
    \caption{Left: microlensing event rate per source star for $3\leq t_E\leq 300$ day events in the VVV survey as we estimate them from the \citet{Husseiniova2021} event list and detection efficiencies, along with the SP-H25 simulated rates with and without the NSD component. Right: source flux fraction distribution of SynthPop-PopSyCLE simulated events for a mock VVV survey, shown in contrast with the uniform distribution used to calculate detection efficiencies.}
    \label{fig:vvv_mulens}
\end{figure}

We limit the SP-H25 simulated events to those with $3 \leq t_E \leq 300$ days to match the $t_E$ range of the VVV detection efficiencies and correct the number of sources using the blending scheme described above with a blend radius of 0.36\arcsec. Figure \ref{fig:vvv_mulens} shows these estimated VVV event rates, as well as SP-H25 event rates with a baseline magnitude limit of Ks$<17$.  In addition to the regular SP-H25 model, we show a version of this simulation that excludes the nuclear stellar disk, which reduces event rates by $\lesssim$10\% at $|b|<1^\circ$. The RGBTDS' Galactic Center field's NSD contribution may be higher as more stars in and behind the NSD will be detectable at \iromann's improved depth.

The SP-H25 simulation over-predicts the event rates by $0-40$\% in the inner few degrees of the Galactic plane, which is in stark contrast with the UKIRT results, where our predictions are consistent with the estimated lower limits and might under-predict true rates.
We note that the \citet{Husseiniova2021} detection efficiencies assume a uniform distribution of the blending parameter, which may result in inaccurate efficiencies if the true distribution of blend fractions is not uniform. Lower source flux fraction events are generally more difficult to detect as the microlensing signal is diluted by light from neighboring stars.
To test this concern, we ran a few SP-H25 model sightlines through the PopSyCLE \citep{Lam2020} event simulator, using \synthpop for the base Galactic model instead of Galaxia and not injecting additional compact objects. 
Selecting events with baseline Ks $<18$, we show the source flux fraction distribution in Figure \ref{fig:vvv_mulens}. These distributions are weighted heavily toward low source flux fraction. Thus, the detection efficiencies used in our observational event rate calculation are likely significantly overestimated for these purposes, which would mean that our observed event rates from estimated from the event list and detection efficiencies of \citet{Husseiniova2021} are significantly underestimated. 

Ultimately, it is difficult to draw firm conclusions from the limited IR microlensing data available. The SP-H25 model agrees with UKIRT's lower limits, thus either agreeing with or under-predicting the true underlying rates. The model over-predicts VVV rates using the detection efficiency correction available, but we show that these detection efficiencies are likely causing the observed event rates to be significantly under-estimated.  We discuss these and other caveats on our model in \S\ref{sec:disc} next.

\section{Discussion} \label{sec:disc}
In Table \ref{tab:tests}, we summarize results of the data-model comparisons described in $\S$\ref{sec:counts}-\ref{sec:mulens}, focused in on just the RGBTDS survey regions (see Figure \ref{fig:test_fields}). The model is more reliable in the lower bulge fields than the Galactic Center field, where the data sets available have more significant uncertainties.

\begin{deluxetable}{p{3cm}p{4.8cm}p{9.2cm}}
\tablecaption{Summary of model-data comparison results, focused exclusively on the portions of the data sets that lie nearest to the \iroman GBTDS fields.    \label{tab:tests}}
\tablehead{ \colhead{Observation category} & 
\colhead{Data source} & \colhead{Model performance}}
    \startdata
    Lower bulge fields & ~ & ~ \\
        ~~Luminosity \& color & WFC3 Bulge Treasury\tablenotemark{a} & 10\% count agreement for most luminosity function bins, worsening ($\lesssim$30\% over-prediction) at the dim end in I, J, and H-band \\
            ~ & OGLE star counts\tablenotemark{b} & Under-predicted star counts by an average of 2\% at I$<$18 and 7\% at I$<$21. \\
            ~ & OGLE\tablenotemark{b} \& VVV\tablenotemark{c} color-magnitude & Red clump too blue by $\sim0.3$ mag. in V-I and well-matched in IR \\
        ~~Kinematics & WFC3 Bulge proper motions\tablenotemark{a} & Bulge-disk separation and dispersions agree within $\sim 1\sigma$ \\
            ~ & Gaia\tablenotemark{d} \& VVV\tablenotemark{c} proper motions\tablenotemark{d} & Good agreement within large observational uncertainties \\
            ~ & BRAVA radial velocities\tablenotemark{d} &  Both mean velocities and mean velocity dispersions are consistent within $\sim1\sigma$.  \\
         ~~Microlensing & Optical surveys\tablenotemark{b} & 60\% event rate per area over-prediction, 30\% event rate per star over-prediction, 20\% optical depth over-prediction, and mean $tE$ consistent within $\sim1\sigma$ \\ 
         ~ & Infrared surveys\tablenotemark{e} & Inconclusive (lack of detection efficiency analysis) \\ 
         \tableline
    Galactic Center field & ~ & ~ \\
         ~~Luminosity \& color & GNS\tablenotemark{f} luminosity functions & Clear red clump in simulation not seen in data, star counts under-predicted in dim K-band and over-predicted in all H- and J-band \\
         ~ & GNS\tablenotemark{f} color-magnitude & Possible underestimation of differential reddening in bulge/NSD stars with limited completeness/quality in the reddest star data \\
         ~ & HST clusters survey\tablenotemark{g} & Over-prediction of star counts by $\sim$40-100\% \\
         ~~Kinematics & HST cluster proper motions\tablenotemark{g} & Well-matched mean proper motions, some over-prediction of velocity dispersions \\
         ~~Microlensing & Infrared surveys\tablenotemark{e} & Inconclusive (lack of detection efficiency analysis)
    \enddata
    \tablenotetext{a}{{\citet{Brown2009,Brown2010}; \url{https://archive.stsci.edu/prepds/wfc3bulge/}}}
    \vspace{-0.1in}
    \tablenotetext{b}{OGLE \citep{Mroz2019}}
    \vspace{-0.1in}
    \tablenotetext{c}{\citet{Minniti2010,Smith2025}}
    \vspace{-0.1in}
    \tablenotetext{d}{{\citet{Howard2008brava,Kunder2012brava}; \url{https://brava.astro.ucla.edu/}}}
    \vspace{-0.1in}
    \tablenotetext{e}{{UKIRT \citep{Wen2023} \& VVV \citep{Husseiniova2021}}}
    \vspace{-0.1in}
    \tablenotetext{f}{GALACTICNUCLEUS Survey \citep{Nogueras-Lara2019}}
    \vspace{-0.1in}
    \tablenotetext{g}{\citet{Hosek2022}}
\end{deluxetable}

\subsection{Stellar Magnitudes, Positions, and Motions}
Overall, SP-H25 matches star counts and CMDs fairly well for typical Galactic bulge fields, including those of the long-running microlensing surveys. 
In the Stanek Window, near the RGBTDS fields, we see an average of 7\% overprediction in J-band counts per bin and 22\% in H-band counts per bin, which would suggest our predictions for \iroman F146 star counts would be off by a value within that range. With good agreement to the bulge-selected population from \citet{Terry2020} in the same field as well, we consider this an improvement in this area over previous models including the Besan{\c c}on model and Galaxia. Within the limited stellar mass range GalMod covered, performance was comparable \citep{Terry2020}, but SynthPop extends to lower stellar masses and thus dimmer stars. 

Color-magnitude diagrams in both the optical and IR show good agreement overall, with some slight deviations such as the RC-smearing issue discussed in \S\ref{sec:ogle_cmd}. Thus, future improvements to the model should include a re-examination of this issue.

With OGLE-IV, we see some deviations on the $\lesssim25$\% level in star counts across the $|b|>1^\circ$ fields. In both OGLE and the Galactic Center-focused data comparisons, we see significant deviations between observed and SP-H25 star counts at low absolute Galactic latitudes ($|b|<1^\circ$). The NSD model used in this work was added with no re-optimization or attenuation of the other populations in that region. Ultimately, much greater care will be necessary in the stellar density distributions in the $|b|\lesssim0.6^\circ$ region. We leave this as well-motivated future work for the continued \synthpop development project.

The proper motion distributions from WFC3, Gaia, and VVV show that the SP-H25 model reproduces kinematics for bulge stars near the Roman fields well, but that near-disk stars may not be well-modeled within a few degrees of the Galactic plane, particularly in optical bands, due to the extinction map used. The choice to use the \citet{Surot2020} extinction map with an exponential disk scaling optimizes the model to bulge stars. For stars at $\lesssim4$ kpc, the line-of-sight dust distribution may cause significant over- or under- extinction, depending on the presence of dust clumps in or beyond the first few kpc out from the Sun. Ultimately, a hybrid dust map which combines a nearby 3-d dust map like that from \citet{Lallement2022} with a Galactic-bulge based map like that from \citet{Surot2020} would be ideal to produce realistic extinction values for all stars. The flexible \synthpop code structure can make this possible as we continue to improve the models over time. Other applications of the \synthpop code seeking to use this model would benefit greatly from selecting an extinction map optimized for their specific region of interest within the Galaxy.

The extinction-related issues are significantly less strong than in near-IR observations than optical. Additionally, the RGBTDS' microlensing survey will focus primarily on bulge stars, which our extinction map captures well. The RGBTDS will focus on small $|l|$, where the gradients we see the model-data disagreement in bulge proper motions have minimal effect. For larger $|l|$ regions, the \iroman Galactic Plane Survey will provide valuable new insights on proper motion distributions and perhaps even radial velocities for the spectroscopic fields.

\subsection{Microlensing Observables}
The SP-H25 model overpredicts microlensing event rates for optical microlensing surveys in the Galactic bulge by $\sim50$\% for event rate per unit area and $\sim20$\% for event rate per source star. A scaling factor determined this way has been used in the past to scale predicted microlensing events in the IR when making projections for the RGBTDS \citep[e.g.][]{Penny2019}, notably including the application of the SP-H24 model for the survey optimization adopted by the Roman Observations Time Allocation Committee \citep{rotac_report}. However, given the complex array of factors that can contribute to this offset, it is not obvious that the optical event rate over-prediction should match the IR event rate over-prediction. In fact, the comparisons with UKIRT data suggest that the H- and K-band event rates are, if anything, underpredicted. A reanalysis of the VVV survey data with additional consideration given to blending in the calculation of detection efficiencies and/or detection efficiency-corrected UKIRT rates would be needed to better answer this question with existing data.


\subsection{SynthPop Framework Limitations}
As described by \citetalias{synthpopI} (\citeyear{synthpopI}), the SynthPop code framework includes some fundamental simplifications for computational feasibility and modular flexibility, which limit our accuracy. A key limitation relevant here is that the \synthpop version used in this work (v1.1.1) does not include stellar multiplicity. This option has since been introduced and will be formalized in the upcoming v2 release.

In \synthpopp, extinction is provided at a reference wavelength by an extinction map for each star's position, and it is scaled using an extinction law for each filter, given a filter's Vega flux-scaled effective wavelength. In reality, extinction in a given filter depends on each star's individual spectrum. This means that for wide bandpasses, extinction will be applied with varying accuracy for different stellar types. This is of particular concern for RGBTDS simulations, as its primary filter will be the F146 band, a very wide filter encompassing Y, J, and H. In our follow-up work applying this model to make updated RGBTDS predictions, we include a correction for this and discuss the impact on prior yield estimates.

\section{Conclusion} \label{sec:concl}
We have presented a Galactic model implementation via \synthpop that we abbreviate as as SP-H25. We described the prior Galactic modeling codes and observational results that shaped this updated model, noting choices that were made to optimize its performance for microlensing predictions and analysis for the upcoming \iroman Galactic Bulge Time Domain Survey. The code used to generate the model-data comparisons in this work is publicly available\footnote{\url{https://github.com/synthpop-galaxy/synthpop-model-eval}}.

The model and data agree well in star counts and kinematics across much of the Galactic bulge, but show some significant deviations within $|b|<1^\circ$. The model over-predicts observed optical microlensing event rates toward the bulge, but comparisons to limited IR survey event rates are inconclusive.
Given the model's performance against observational data, the predictions it can provide for the contiguous lower bulge fields in the RGBTDS are expected to be more reliable than those for the Galactic Center field. In Paper II in this pair, we run SP-H25 simulations for the RGBTDS and explore the implications for its science yields.

Thus far, microlensing simulations have not included a detailed examination of additional Galactic components including the Nuclear Stellar Disk (NSD) and the Nuclear Star Cluster (NSC). The {\tt genstars} Galactic model includes an NSD adopted from \citet{Sormani2022}, the same model which we include in SP-H25. However, the SP-H25's Galactic model performance is still poor at $|b|\lesssim0.5^\circ$, and {\tt genstars}' difference in performance has not been examined in a publication. An NSC has not been included in any such simulations. Thus, an improved handling of these additional Galactic populations near the GC will be necessary to accurately predict and interpret the RGBTDS' GC field results. 

With \synthpopp's flexible model design, we can continue to improve these components, and both \iromann's Galactic Bulge Time Domain Survey and Galactic Plane Survey will provide incredibly valuable data sets for this analysis. Version 2 of the \synthpop framework is nearly ready for public release and will remove some inefficiencies, assumptions, and inaccuracies of the prior versions presented in this work, providing an improved platform on which to perform a model re-optimization with new data sets like those from \iroman and the upcoming Gaia Data Release 4.

\begin{acknowledgments}
The authors thank Naoki Koshimoto for helpful discussions in implementing the \citet{Koshimoto2021model} model and \citet{Sormani2022} nuclear stellar disk into {\sc SynthPop}. We thank Dylan Paterson who discovered a bug in the bulge kinematics in a previous version of this work when studying differences between various \synthpop models and VVV proper motion data.
MJH thanks their PhD thesis committee (advisor Jason Wright, Ian Czekala, Caryl Gronwall, Kevin Luhman, and Sarah Shandera) for useful discussions around a preliminary version of this work.

MJH's work was supported by the Penn State Extraterrestrial Intelligence Center, the Pennsylvania Space Grant Consortium and the Heising-Simons Foundation under grant No. 2022-3542. 
ALC was supported by NASA award 80NSSC24M0022 and conducted portions of this research on the Unity Cluster of the College of Arts and Sciences at the Ohio State University. 
MN acknowledges support from NASA award 80NSSC24K0881. 
MTP was supported by NASA awards 80NSSC24M0022 and 80NSSC24K0881, and portions of this research were conducted with high performance computing resources provided by Louisiana State University (\url{http://www.hpc.lsu.edu}).
PM acknowledges that this work was partly performed under the auspices of the U.S. Department of Energy by Lawrence Livermore National Laboratory under Contract DE-AC52-07NA27344. The document number is LLNL-JRNL-XXXXX. This work was supported by the LLNL LDRD Program under Project 22-ERD-037.

Funding for the Roman Galactic Exoplanet Survey Project Infrastructure Team is provided by the Nancy Grace Roman Space Telescope Project through the National Aeronautics and Space Administration grant 80NSSC24M0022, by The Ohio State University through the Thomas Jefferson Chair for Space Exploration endowment, and by the Vanderbilt Initiative in Data-intensive Astrophysics (VIDA). NSA acknowledges support from the National Science Foundation under grant No. 1909641, the Heising-Simons Foundation under grant No. 2022-3542, and the H2H8 foundation.

This research has made use of NASA’s Astrophysics Data System. 
This work has made use of data from the European Space Agency (ESA) mission
{\it Gaia} (\url{https://www.cosmos.esa.int/gaia}), processed by the {\it Gaia}
Data Processing and Analysis Consortium (DPAC,
\url{https://www.cosmos.esa.int/web/gaia/dpac/consortium}). Funding for the DPAC
has been provided by national institutions, in particular the institutions
participating in the {\it Gaia} Multilateral Agreement.

\end{acknowledgments}

\vspace{5mm}

\software{\synthpop (\citetalias{synthpopI}~\citeyear{synthpopI}, \citealp{huston_2026_20334473}), astropy \citep{astropy:2013,astropy:2018,astropy:2022}, astroquery \citep{astroquery}, h5py \citep{andrew_collette_2023_7560547}, numpy \citep{harris2020array}, pandas \citep{mckinney-proc-scipy-2010,reback2020pandas}, scipy \citep{2020SciPy-NMeth}}

\bibliography{refs}{}
\bibliographystyle{aasjournalv7}

\appendix 
\section{Microlensing Statistic Calculation Methods} \label{app:weights}
Microlensing optical depths, event rates, and timescales can be simulated in several ways with varying levels of complexity. \citet{Paczynski1986} presented an integral expression for microlensing optical depth at a particular source distance:
\begin{equation}
    \tau = \int_0^{D_{s}} \frac{4\pi G D_l (D_s-D_l)}{D_s c^2} \rho(D_l) dD_l, 
\end{equation}
where $D_s$ is the distance to a source, $D_l$ is the distance to the lens, and $\rho(D_l)$ is the mean mass density of lens objects at the distance $D_l$. We note that some of these original papers use $D_d$ instead of $D_l$, referring to the lens as a `deflector,' but we use $D_l$ throughout for consistency.
This equation assumes that all sources are at the same distance, and a different source distance distribution would require a more complex integral that essentially averages the optical depth computed in this way across a realistic distribution of source distances.
If we maintain the single source distance assumption for now and assume also that all lenses have the same mass, this becomes:
\begin{equation}
    \tau = \int_0^{D_s} \frac{4\pi G M_l D_l (D_s-D_l)}{D_s c^2} n(D_l) dD_l = \int_0^{D_s} \pi \theta_E^2 n(D_l) D_l^2 dD_l,
\end{equation}
where $n(D_l)$ is the lens number density at the lens distance.

We can introduce additional complexity (removing the assumption of equal mass lenses and a single source distance) when performing Monte Carlo integration over a simulated star catalog from a Galactic model:
\begin{equation}
    \tau = \frac{\pi}{N_s \Omega_L} \sum^{N_s}  \sum^{N_l}_{D_s>D_l} \theta_E^2.
\end{equation}
This equation calculates the fraction of area on-sky covered by Einstein rings for an average source star. 
In both integral and summation forms, the key value that we are evaluating for every potential microlensing event is $\pi \theta_E^2$, the angular area of each Einstein ring. When moving from the integral form to the summation form, the summation over each valid lens star corresponds to the integration over $n(D_l)\Omega_l D_l^2dDl$, which is the total number of lens stars. In the summation, our lens catalog corresponds to a finite region of sky, $\Omega_l$, which we must divide our values by to convert from area on-sky to fractional sky coverage. However, in the integral form, this $1/\Omega_l$ factor cancels with the $\Omega_l$ in the volume integration element, and thus neither are present in the integral formulations above. The integral form assumes a single source distance, while the summation form averages over all catalog sources. Our Monte Carlo integration equation assumes that the lens and source stars are draw from Galactic model catalogs across a finite solid angle, establishing a cone-shaped volume region on sky. 

\citet{Paczynski1996} showed that the number of events $N$ expected during a time interval $\Delta t$, assuming that all events have identical timescales is:
\begin{equation}
    N = \frac{2 N_s \tau \Delta t}{\pi t_E}.
\end{equation}
The translates to an event rate per star $\Gamma_*$ of:
\begin{equation}
    \Gamma_* = \frac{N}{N_s \Delta t} = \frac{2 \tau}{\pi t_E}.
\end{equation}
\citet{Mao2008} showed that the integral form of the total event rate $\Gamma$ is:
\begin{equation}
    \Gamma_* = \int_0^{D_s} \frac{2\theta_E^2}{t_E} n(D_d) D_d^2 dD_d.
\end{equation}

When performing Monte Carlo integration from a catalog, we can again re-introduce variation in the timescales and Einstein ring radii:
\begin{equation}
    \Gamma_* = \frac{1}{N_s \Omega_l} \sum^{N_s} \sum^{N_l}_{D_s>D_l} 2\mu_{\rm rel}\theta_E.
\end{equation}
This equation calculates the fractional rate of angular area on-sky swept out by Einstein rings for an average source star. 
Removing the assumption of a uniform timescale, we instead use the average timescale $\overline{t_E}$ to relate optical depth and event rate:
\begin{equation}
    \Gamma_* = \frac{2 \tau}{\pi \overline{t_E}}.
\end{equation}
This can be converted to an event rate per unit solid angle instead of per source star by simply multiplying it by the number of source stars per unit solid angle:
\begin{equation}
    \Gamma_\Omega = \frac{N_s}{\Omega_s} \Gamma_*.
\end{equation}
The translation between integral and summation forms here works like the optical depth calculation, with the rate of area swept out by each potential event's Einstein ring being the key value from each event instead of the instantaneous angular Einstein ring area.

In Monte Carlo integration, we calculate the average timescale as:
\begin{equation}
    \overline{t_E} = \frac{\sum^{N_s} \sum_{D_s>D_l}^{N_l} t_E * \theta_E \mu_{rel}}{\sum^{N_s} \sum_{D_s>D_l}^{N_l} \theta_E \mu_{rel}},
\end{equation}
a weighted average where each lens-source pair's timescale is weighted by a value proportional to its contribution to the overall event rate, $\mu_{\rm rel} \theta_E$. 
This equation is consistent with the relationship between event rate and optical depth and their summation forms as defined above. 

Some microlensing simulators e.g. {\tt genulens} \citep{Koshimoto2022} take an approach in between these two methods, where points are sampled along a line in distance rather than a conical volume, which requires the inclusion of a $D_l^2$ factor in the weights used in summation, as the volume included in a distance element $dD_l$ for constant solid angle $\Omega_l$ increases as $D_l^2$. 

Past Manchester-Besan{\c c}on Microlensing Simulator (MaB$\mu$lS) simulations \citep{Kerins2009,Awiphan2016,Specht2020} have applied the additional $D_l^2$ weighting factor in the computation of the average $t_E$ when sampling from cone-volume catalogs.
This results in the double counting the $D_l^2$ factor, skewing $t_E$ distributions incorrectly to favor nearby lenses.
We tested how this change in weighting would impact simulation results in this case by computing the timescale distribution using three different methods. 

First, we generated large SP-H25 model catalogs with no magnitude cut for several sample OGLE fields. We ran them through a resolved microlensing simulation via PopSyCLE \citep{Lam2020,Rose2022,Abrams2025}, using these catalogs as the base model instead of Galaxia and performing no additional injection of compact objects or binary companions. PopSyCLE propagates stellar motions across the sky with time and checks for close encounters on some user-determined timescale to find microlensing events for some user-determined threshold impact parameter $u_0$. We used a 10-day cadence (which \citealp{Lam2020} showed is reliable for events with $t_E>3$ days) and keep only $u_0\leq1$ events to be consistent with the Monte Carlo integration method. Next, we applied the Monte Carlo integration approach to the same catalogs using both weighting schemes. 

\begin{figure}
    \centering
    \includegraphics[width=\linewidth]{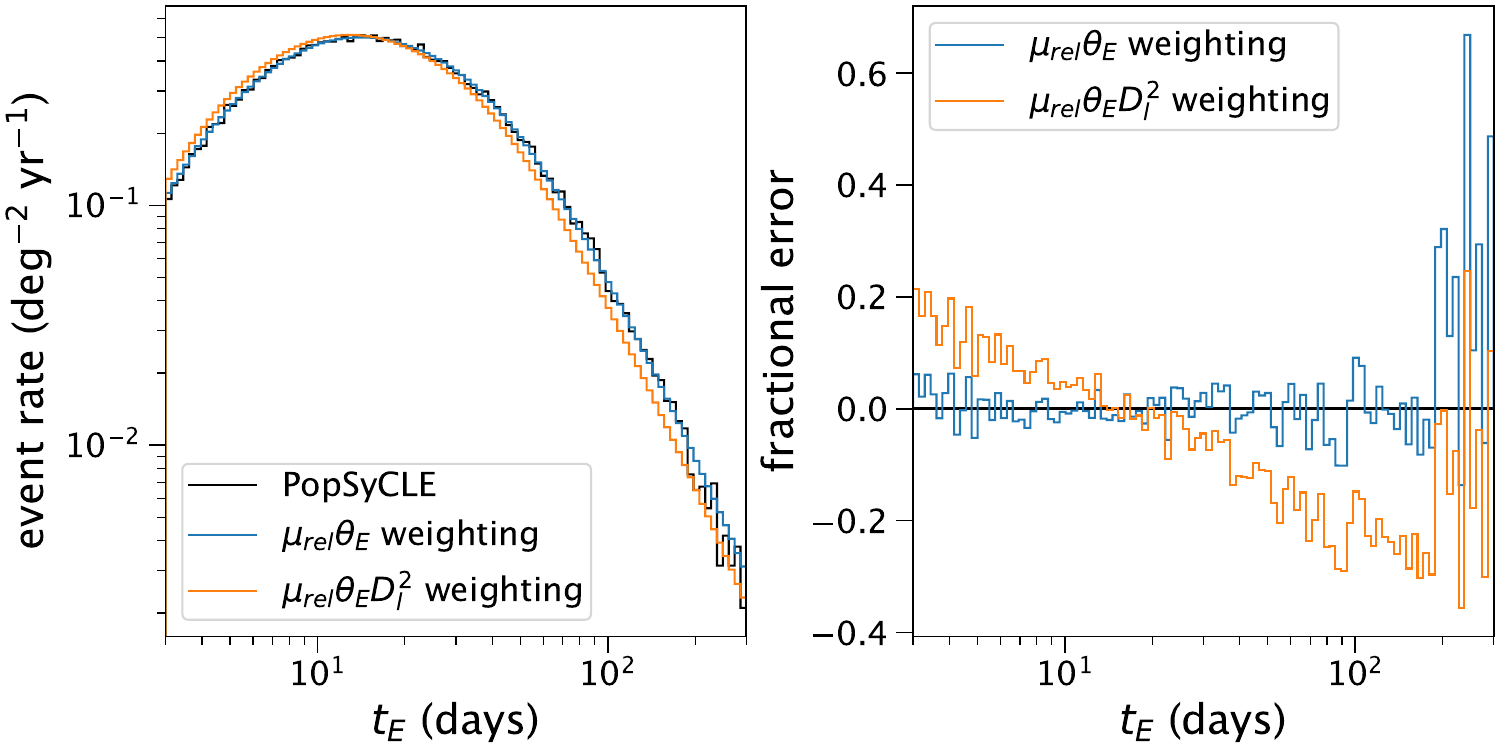}
    \caption{Left: microlensing event timescale ($t_E$) distributions for each selection/weighting method. Right: fractional error of each Monte Carlo integration weighting scheme's resulting $t_E$ distribution relative to the PopSyCLE result.}
    \label{fig:avg_te}
\end{figure}

The resulting $t_E$ distributions for each approach are shown in Figure \ref{fig:avg_te}, as well as the fractional error of each weighting scheme's result against that from PopSyCLE. 
The $\mu_{\rm rel}\theta_E$ scheme matches well until $t_E>200$ days, where event counts are extremely low, leading to poor statistics in the PopSyCLE method. The $\mu_{\rm rel}\theta_E D_l^2$ weighting scheme produced a result that is skewed toward shorter events. Overall, we get an average timescale for events limited to $3 \leq t_E \leq 300$ days of 22.33 days from PopSyCLE, 22.32 days from the $\mu_{\rm rel}\theta_E$-weighted statistical method, and 20.54 from the $\mu_{\rm rel}\theta_E D_l^2$-weighted statistical method. Thus, using the $\mu_{\rm rel}\theta_E D_l^2$ weights may result in an under-estimated $\overline{t_E}$. When this is combined with $\tau$ to estimate event rates, event rates may be over-estimated. In this particular example, the event rate overestimation is $\sim$10\%, though the precise difference for any given application would be sight line and magnitude limit dependent.

The MaB$\mu$lS results presented by \citet{Awiphan2016,Specht2020} included an extension of the Besan{\c c}on model's IMF into the brown dwarf regime that was tuned to reproduce observed average timescales. This tuning compensated for the extra $D_l^2$ weighting factor and allowed the simulator to produce results that agreed well with observation. 

\end{document}